\documentclass[a4paper, 14pt, conference,onecolumn]{ieeeconf}      

\IEEEoverridecommandlockouts                              

\usepackage{graphicx}      
\usepackage{amsmath}
\usepackage{fancyhdr}
\usepackage[english]{babel}
\usepackage[latin1]{inputenc}   
\usepackage{amsfonts,amsmath,bm,hhline}
\usepackage{epsfig}
\usepackage{graphicx}
\usepackage{color}
\usepackage[hang]{caption2}
\usepackage[normalsize,it,hang]{subfigure}
  
\title{\LARGE \bf Defense against DoS and load altering attacks via model-free control: \\ A proposal for a new cybersecurity setting
}

\author{Michel Fliess$^{1,3}$, C\'edric Join$^{2,3}$ and Dominique Sauter $^{2}$
\thanks{$^{1}${LIX (CNRS, UMR 7161), \'Ecole polytechnique, 91128 Palaiseau, France {\tt Michel.Fliess@polytechnique.edu}}}
\thanks{$^{2}${CRAN (CNRS, UMR 7039)), Universit\'{e} de Lorraine, BP 239, 54506 Vand{\oe}uvre-l\`{e}s-Nancy, France \newline {\tt \{cedric.join, dominique.sauter\}@univ-lorraine.fr}}}
\thanks{$^{3}${AL.I.E.N., 7 rue Maurice Barr\`{e}s, 54330 V\'{e}zelise, France \newline {\tt \{michel.fliess, cedric.join\}@alien-sas.com}}}
}

\begin{document}

\maketitle
\thispagestyle{empty}
\pagestyle{empty}

\begin{abstract}
Defense against cyberattacks is an emerging topic related to fault-tolerant control.  In order to avoid difficult mathematical modeling, model-free control (MFC)  is suggested as an alternative to classical control.  For illustration purpose a Load Frequency Control of multi-areas power network is considered. In the simulations, load altering attacks and Denial of Service (DoS) in the communication  network  are applied to the system.  Our aim is to compare the impact of cyberattacks on control loops closed via respectively a classical controller in such situations and a model-free one.  Computer experiments show  impressive results with MFC.



\end{abstract}

\section{Introduction}

The design of secure and safe \emph{Networked Control System} (\emph{NCS}) is of high importance in the control of large-scale critical infrastructures or industrial plants such as power grids, transportation systems, communication networks, oil and gas pipelines, water distribution or waste-water treatment systems and irrigation networks \cite{c1}.  Using ``open'' public and also wireless networks for the communication within NCS can generate severe security problems since an unauthorized access (``cyberattack'') is possible in the control system. The security of control systems against malicious attacks has received a great deal of attention over the past few years, in particular after the Stuxnet attack against Iran nuclear installations in 2010 \cite{c2}. Specific analysis tools as well as monitoring mechanisms have been developed to enforce system security and reliability \cite{c3,c4}. Information security approach may provide some protection methods that help in improving the security of control systems, but these methods appear to be not sufficient for the defense of the systems against malicious attacks able to bypass information security layers, as in the case of Stuxnet incident in 2010. As pointed out in \cite{c5} the security of \emph{Cyber-Physical Systems} (\emph{CPS}) integrating computation, communication and physical capabilities must be improved from both information technology and control theory. 
The \emph{cyber-physical attacks} (\emph{CPA}), on both the physical layer and the cyber layer, are modeled as additive signals of short duration on both system equations. Attackers can break into the communication channels, enabling them to modify the command signals, control signals or sensor measurements for disrupting the systems. CPA in CPS, summarized in \cite{c5,c6,c7}, may be divided into several categories: 
\begin{itemize}
\item \emph{Denial of Service} (\emph{DoS}) attacks in \cite{c8,c9}: adversaries aim at disrupting temporarily or indefinitely the exchange of data among entities in the network. 
\item\emph{Integrity} or \emph{Man in The Middle} (\emph{MITM}) attacks in \cite{c10,c11,c12,c13,c14}: adversaries inject false data on control signals or on information transmitted by sensors to the plant via communication channels, and finally physical attacks on sensors and actuators close to faults. 
\item \emph{Replay attack} in \cite{c15} can be viewed as a deception attack on control signals coordinated with the generation of artificial delays on measurements. 
\end{itemize}
Among such a huge number possibilities (see also \cite{review,survey,md0,zhao}), we concentrate here on Denial of Service and  \emph{load altering} \emph{attacks} which correspond to a large body of concrete situations. The defense against those strikes is connected to a classic topic in control engineering, namely \emph{fault-tolerant control} (see, \textit{e.g.}, \cite{noura2,blanke}), \textit{i.e.}, a set of techniques for mitigating the effects of the unavoidable faults which occur in any control system.

\emph{Model-free control}, or \emph{MFC}, in the sense of \cite{mfc13,nicu} is chosen for the following reasons:
\begin{itemize}
\item It has been already successfully applied many times (see, \textit{e.g.}, \cite{wang} for an automated vehicle) including in fault-accommodation \cite{toulon,park}.
\item Most of the existing defense approaches rely on a mathematical modeling (see, \textit{e.g.}, \cite{c25,c26,c27,c28,md} for power systems) which too often is most difficult to derive. A model-free setting might therefore be fruitful (see, \textit{e.g.}, \cite{kemal,cui,qiu}).
\item Load altering attacks ought to be related to actuator's faults. It has been proven \cite{mfc13}, \cite{toulon} that model-free control is quite efficient in actuator's fault accommodation.
\item Some DoS attacks might look as \emph{packet losses} (see, \textit{e.g.}, \cite{loss}). It has been shown \cite{iot} via numerical and concrete experiments that model-free control exhibits excellent performances in spite of severe losses. 
\end{itemize}

Our proposal is illustrated by several computer experiments. They are based on \cite{ex1,ex2} where 
\begin{itemize}
\item the application of network technology in the power grid makes the \emph{load frequency control} (\emph{LFC}) system more vulnerable to various kinds of attacks (see, \textit{e.g.}, \cite{bevrani} for a general presentation);
\item DoS and load alteration attacks fit very well.
\end{itemize}

Our paper is organized as follows. Section \ref{mfc} reviews MFC which is certainly unknown to most of the experts in cybersecurity. In particular Section \ref{accom}  explains how to react against load altering attacks. Computer experiments are displayed in Section \ref{exp} with many Figures and $2$ Tables, which show the efficiency of our approach. Some concluding remarks may be found in Section \ref{conclusion}.

\section{What is model-free control?}\label{mfc}
\subsection{Ultra-local model}
The unknown global description of the plant is replaced by the following first-order \emph{ultra-local model}:
\begin{equation}
\dot{y} = F + \alpha u \label{1}
\end{equation}
where:
\begin{enumerate}
\item The control and output variables are respectively $u$ and $y$.
\item $\alpha \in \mathbb{R}$ is chosen by the practitioner such that the three terms in Equation \eqref{1} have the same magnitude.
\end{enumerate}
The following comments are useful:
\begin{itemize}
\item $F$ is \emph{data driven}: it is given by the past values of $u$ and $y$.
\item $F$ includes not only the unknown structure of the system but also any disturbance.
\end{itemize}




\subsection{Intelligent controllers}
Close the loop with the \emph{intelligent proportional controller}, or \emph{iP}, 
\begin{equation}\label{IP}
u = - \frac{F_{\text{est}} - \dot{y}^\ast + K_P e}{\alpha}
\end{equation}
where
\begin{itemize}
\item $y^\ast$ is the reference trajectory,
\item $e = y - y^\star$ is the tracking error,
\item $F_{\text{est}}$ is an estimated value of $F$
\item $K_P \in \mathbb{R}$ is a gain.
\end{itemize}
Equations \eqref{1} and \eqref{IP} yield
\begin{equation*}
	\dot{e} + K_P e = F - F_{\text{est}}
			\label{equa15}
	\end{equation*}
If the estimate $F_{\text{est}}$ is ``good'': $F - F_{\text{est}}$ is ``small'', \textit{i.e.}, $F - F_{\text{est}} \simeq 0$,  then $\lim_{t \to +\infty} e(t) \simeq 0$ if $K_P > 0$. It implies that the tuning of $K_P$ is straightforward. This is a major difference with the tuning of ``classic'' PIDs (see, \textit{e.g.}, \cite{astrom}).

\subsection{Estimation of $F$}\label{F}
A real-time estimate of $F$ is given by (see \cite{mfc13} for more details)
\begin{equation}\label{int}
{\small F_{\text{est}}(t)  =-\frac{6}{\tau^3}\int_{t-\tau}^t \left\lbrack (\tau -2\sigma)y(\sigma)+\alpha\sigma(\tau -\sigma)u(\sigma) \right\rbrack d\sigma }
\end{equation}
where $\tau > 0$ is ``small.'' This integral, which is a low pass filter, may of course be replaced in practice by a classic digital filter.

\subsection{MIMO systems}
Consider a multi-input multi-output (MIMO) system with $m$ control variables $u_i$ and $m$ output variables $y_i$, $i = 1, \dots, m$, $m \geq 2$. It has been observed in \cite{toulon} and confirmed by all encountered concrete case-studies (see, \textit{e.g.}, \cite{pmsm}), that such a system may usually be regulated via $m$ monovariable ultra-local models:
\begin{equation*}\label{multi}
\dot{y}_{i} = F_i + \alpha_i u_i
\end{equation*}
where $F_i$ may also depend on $u_j$, $y_j$, and their derivatives, $j \neq i$.

\subsection{Actuator's fault accommodation}\label{accom}
We assume that cyberattacks can be represented as additive signals applied to the controller output.  In Equation \eqref{1} write the input variable 
$$u = u_{\rm attack} + v$$ 
where $u_{\rm attack}$ (resp. $v$) is an unwanted (resp. the desired) quantity. It yields 
$$\dot{y} = \frak{F} + \alpha v$$ 
where 
$$\frak{F} = F + \alpha u_{\rm attack}$$ 
It is straightforward to adapt the iP \eqref{IP} 
\begin{equation*}\label{ip}
v = - \frac{ \frak{F}_{\text{est}}- \dot{y}^\ast + K_P e}{\alpha}
\end{equation*}
and the estimate $\frak{F}_{\text{est}}$ of $\frak{F}$ in Formula \eqref{int}
\begin{equation*}\label{integral}
{\small {\frak F}_{\text{est}}(t)  =-\frac{6}{\tau^3}\int_{t-\tau}^t \left\lbrack (\tau -2\sigma)y(\sigma)+\alpha\sigma(\tau -\sigma)v(\sigma) \right\rbrack d\sigma }
\end{equation*}

\section{Application to power network}\label{exp}

In power systems, LFC used for frequency stabilization \cite{c16,c17} is one of the most essential operational functions. Considering interconnected generation/distribution systems the main objective of LFC is to ensure the balance between load and generation in each control area \cite{c18,c19,c20}. However, the LFC system in modern power systems tends to use open communication networks to transmit control/measurement signals, thus making the LFC system more vulnerable to cyberattacks such as denial of service (DOS) attacks. 

Defense against cyberattacks is an emerging topic for power transmission and distribution systems \cite{c21,c22}. While many works focus on detection \cite{c23,c24} and isolation of the attack signal, few of them actually consider the design of a complete defense mechanism, which is essential for the robustness of the control system under attack.

To date, many researchers have applied a significant effort in favor of the LFC system regarding the defense against DoS attacks. Considering DoS attacks as network-induced sent disturbances, some robust LFC systems for interconnected power systems have been developed as for example in \cite{c25,c26,c27,c28}. Nevertheless, most of these approaches rely on the knowledge of a model of the LFC system. More recently another \emph{model-free} approach, inspired by \emph{fault tolerant control} (\emph{FTC}) \cite{c29}, has  been proposed  \cite{c30,c31}.

\subsection{Power grid}
\begin{figure}[!ht]
\centering%
{\epsfig{figure=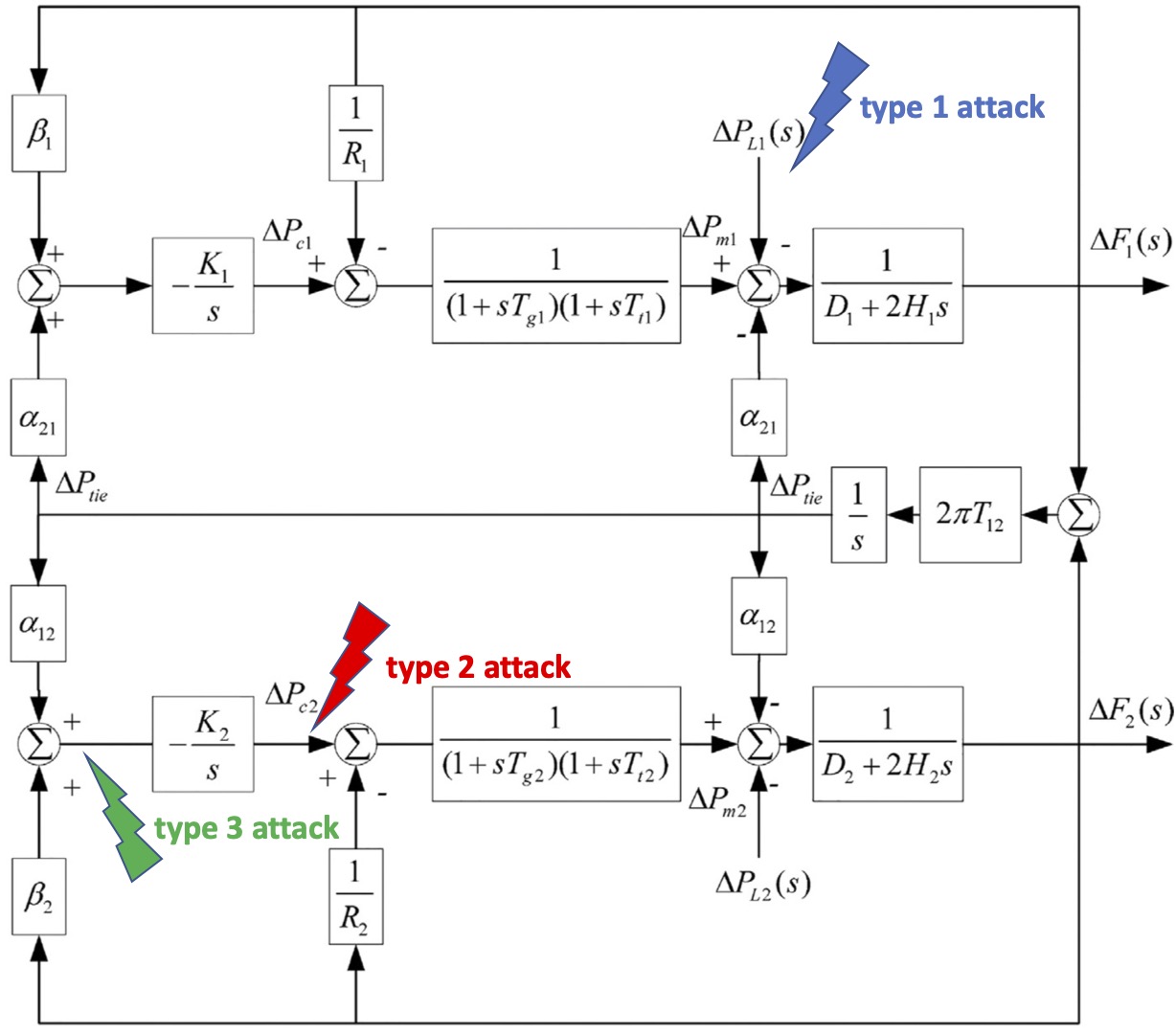,width=.46\textwidth}}
\caption{Block diagram of a power system}\label{scheme} \end{figure}

A large power system consists of a number of interconnected control areas, which are connected by tie lines. The LFC is used to maintain the system frequency and power exchange between the tie lines at a predefined value (see, \textit{e.g.}, \cite{bevrani}). As illustrated in Figure 1, the typical components of an LFC system installation are the controller, turbine, generator and load.  The input to the controller is the \emph{Area Control Error}, or \emph{ACE}. The ACE is defined for each zone as a linear combination of the total power exchanged and the frequency deviations from the respective programmed and nominal values. The LFC system relies on the communication between the sensors and the \emph{Energy Management System}, or \emph{EMS}, and is therefore exposed to high risks of cyber-intrusion. 
In the device displayed in Figure \ref{scheme} load variations are mitigated via phase shift minimization by two generators (see \cite{ex1,ex2} for more details). Note moreover that power grids were already investigated via model-free control \cite{sarajevo,ferrari}.

\begin{figure}[!ht]
\centering%
{\epsfig{figure=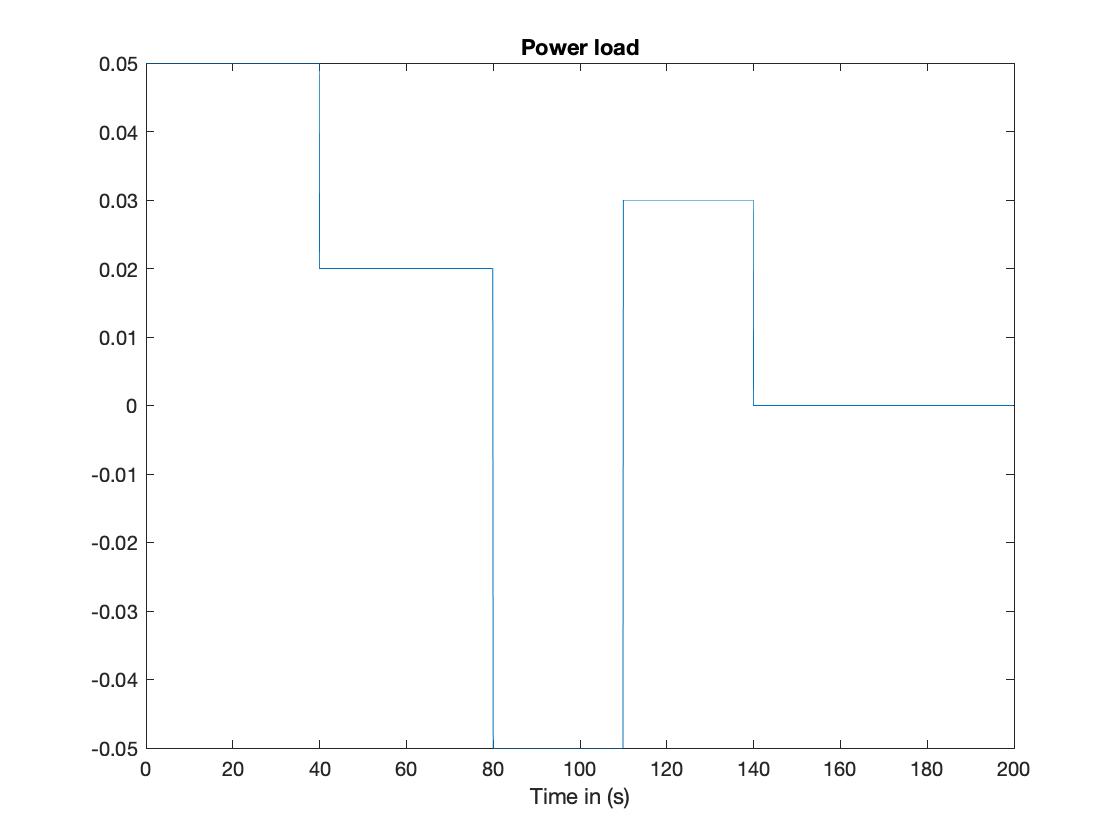,width=.46\textwidth}}
\caption{Nominal load variations}\label{Load}
\end{figure}

Three  types of attacks are considered:
\begin{enumerate}
\item {\bf Type 1 attack (blue)}: very important additive load variation of short time (load altering attack).
\item {\bf Type 2 attack (red)}:  blocking the control to generator 2 (DoS attack).
\item {\bf Type 3 attack (green)}:  blocking the measures of generator 2 (DoS attack).
\end{enumerate}
The pure integrators $\frac{K_1}{s}$, $\frac{K_2}{s}$ in Figure \ref{scheme} are replaced by two model-free controllers defined by Equations \eqref{1}-\eqref{IP} in order to insure a better defense. See \cite{alinea}, which is devoted to traffic regulation on motorways, for a similar result.

\subsection{Various scenarios}
In all our computer simulations, $K_1 = K_2 = 1$, and $\alpha = 10$, $K_P = 0.3$ in Formula \eqref{IP}. Pure integrators are compared to model-free control. The load variations, which are depicted in Figure \ref{Load}, are identical. 

\begin{table*}[!h]
\begin{center}
\caption{Sum of tracking errors for 2 lines}\label{tb1}
\begin{tabular}{|c||c|c|c|c|}
\hline
Control types/scenarios& $\Sigma_t |e_1|$ & $\Sigma_t e_1^2$ & $\Sigma_t |e_2|$ & $\Sigma_t e_2^2$\\ \hline\hline
Integrator/scenario 1 &$9.1106 $& $0.3763$& $2.8981$& $0.0283$ \\ \hline
MFC/scenario 1 & $6.7965$& $0.2832$ & $1.7452$ &$0.0118$ \\ \hline\hline
Integrator/scenario 2 & $29.4256$&$4.6806$ &$9.3662$ &$0.3597$ \\ \hline
MFC/scenario 2 &$21.8710$ &$3.4752 $&$5.5665 $&$0.1418 $\\ \hline
\end{tabular}
\end{center}
\end{table*}

\begin{table*}[!t]
\begin{center}
\caption{Sum of tracking errors for 2 lines: Averaging 100 simulations}\label{tb2}
\begin{tabular}{|c||c|c|c|c|}
\hline
Control types/scenarios& $\Sigma_t |e_1|$ & $\Sigma_t e_1^2$ & $\Sigma_t |e_2|$ & $\Sigma_t e_2^2$\\ \hline\hline
Integrator/scenario 3 & $212.0977$ & $1.8926.10^4$& $1.3241.10^3$&$5.8470.10^5$ \\ \hline
MFC/scenario 3 &$7.3609$ &$0.2884$ &$2.5912$ &$0.0232$ \\ \hline\hline
Integrator/scenario 4 &$6.0832.10^9$&$1.2731.10^{19} $&$2.4134.10^{10}$ &$1.7941.10^{20}$\\ \hline
MFC/scenario 4 &$ 9.6472$&$ 0.5017$&$11.1365 $&$2.4834$ \\ \hline
\end{tabular}
\end{center}
\end{table*}

\subsubsection{No attack} 
\begin{figure*}[!ht]
\centering%
\subfigure[\footnotesize Output (--) and reference trajectory (- -) line 1 ]
{\epsfig{figure=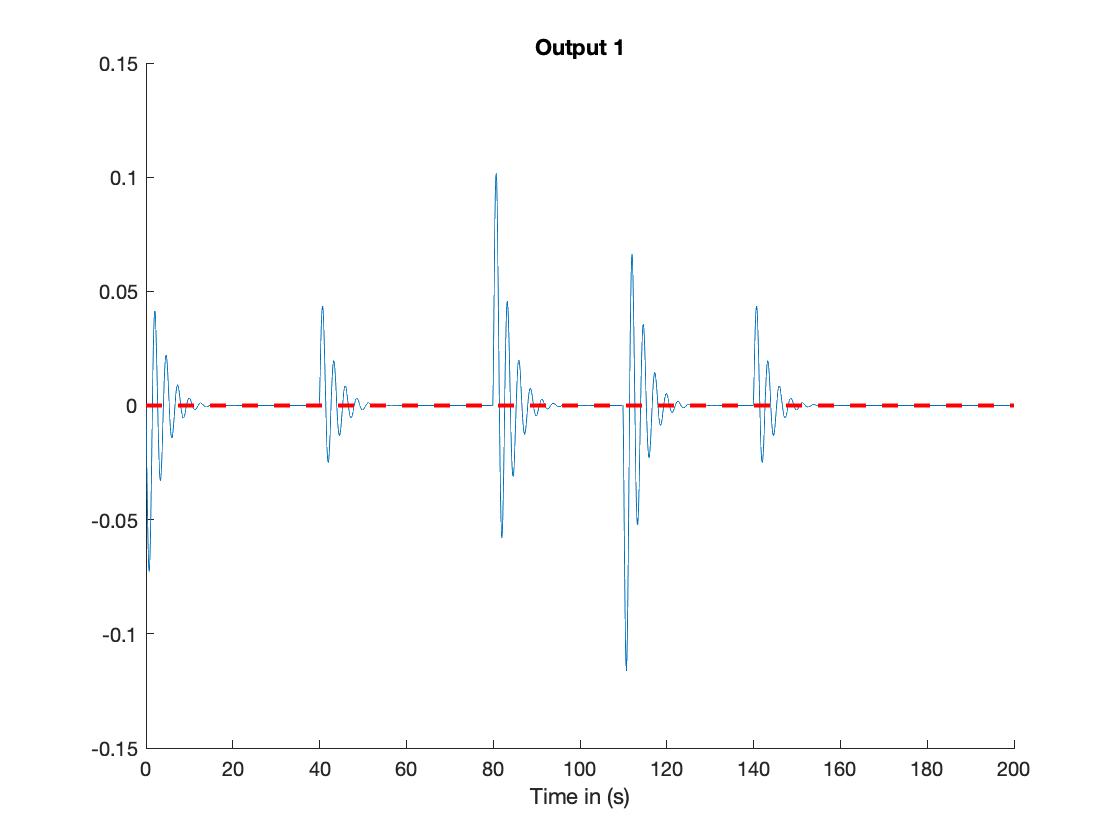,width=.29\textwidth}}
\centering%
\subfigure[\footnotesize Control line 1 ]
{\epsfig{figure=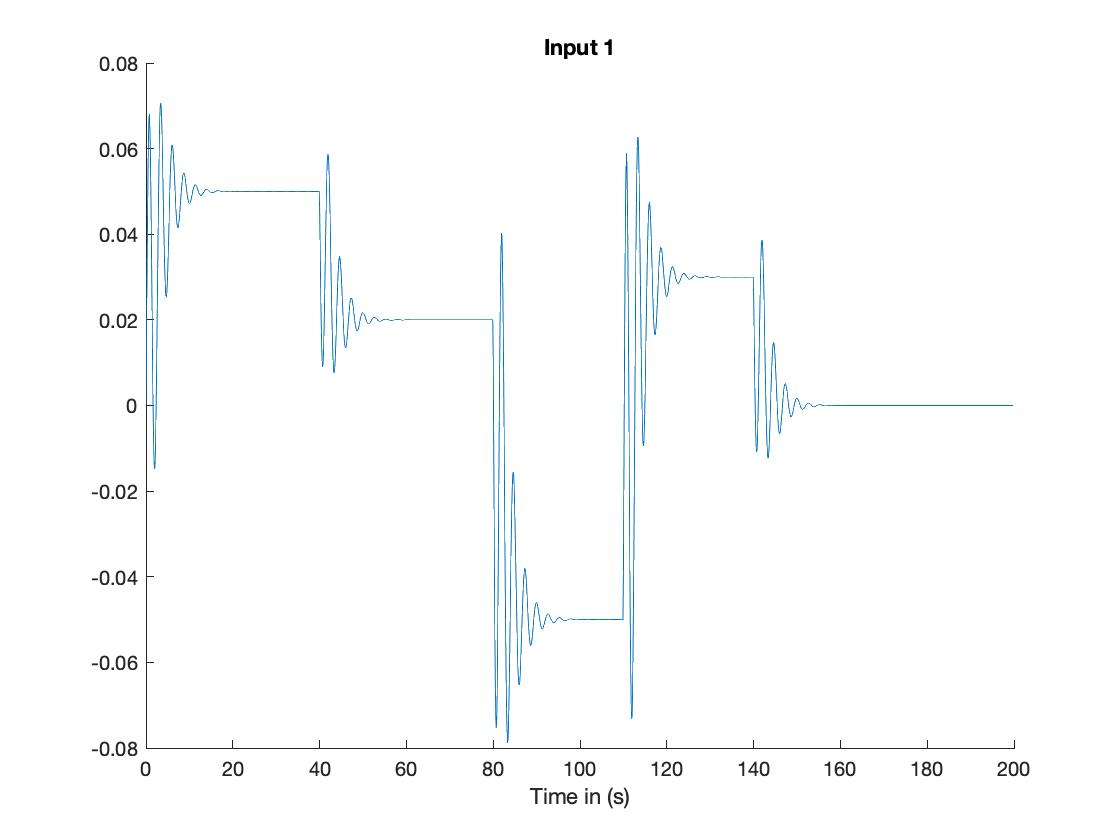,width=.29\textwidth}}
\\
\subfigure[\footnotesize Output (--) and reference trajectory (- -) line 2 ]
{\epsfig{figure=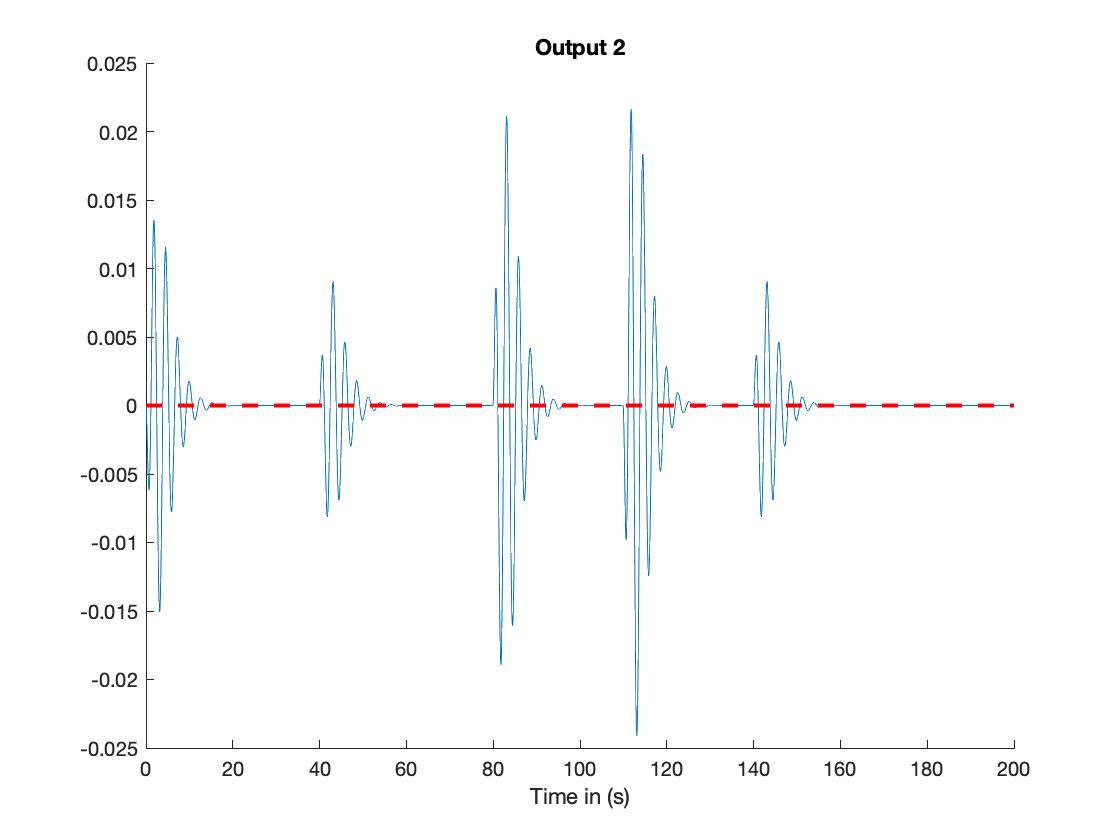,width=.29\textwidth}}
\centering%
\subfigure[\footnotesize Control line 2 ]
{\epsfig{figure=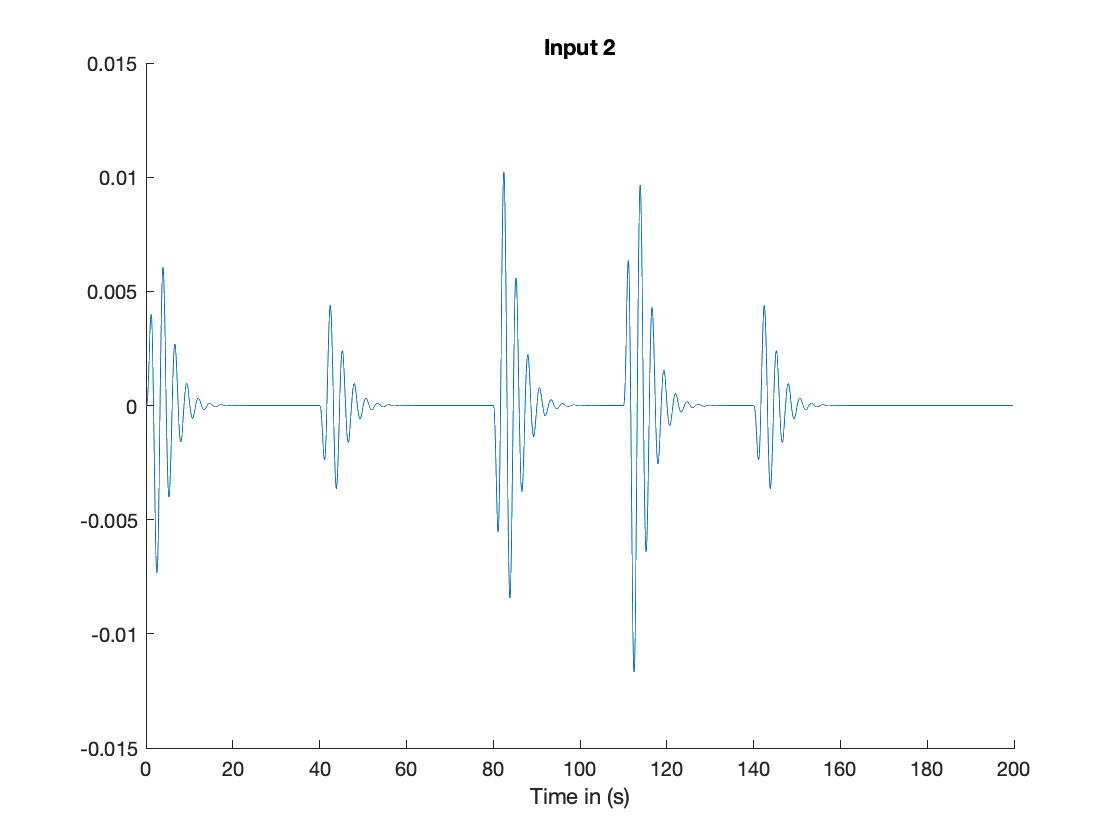,width=.29\textwidth}}
\caption{Integrator: No attack}\label{PID1}
\end{figure*}

\begin{figure*}[!ht]
\centering%
\subfigure[\footnotesize Output (--) and reference trajectory (- -) line 1 ]
{\epsfig{figure=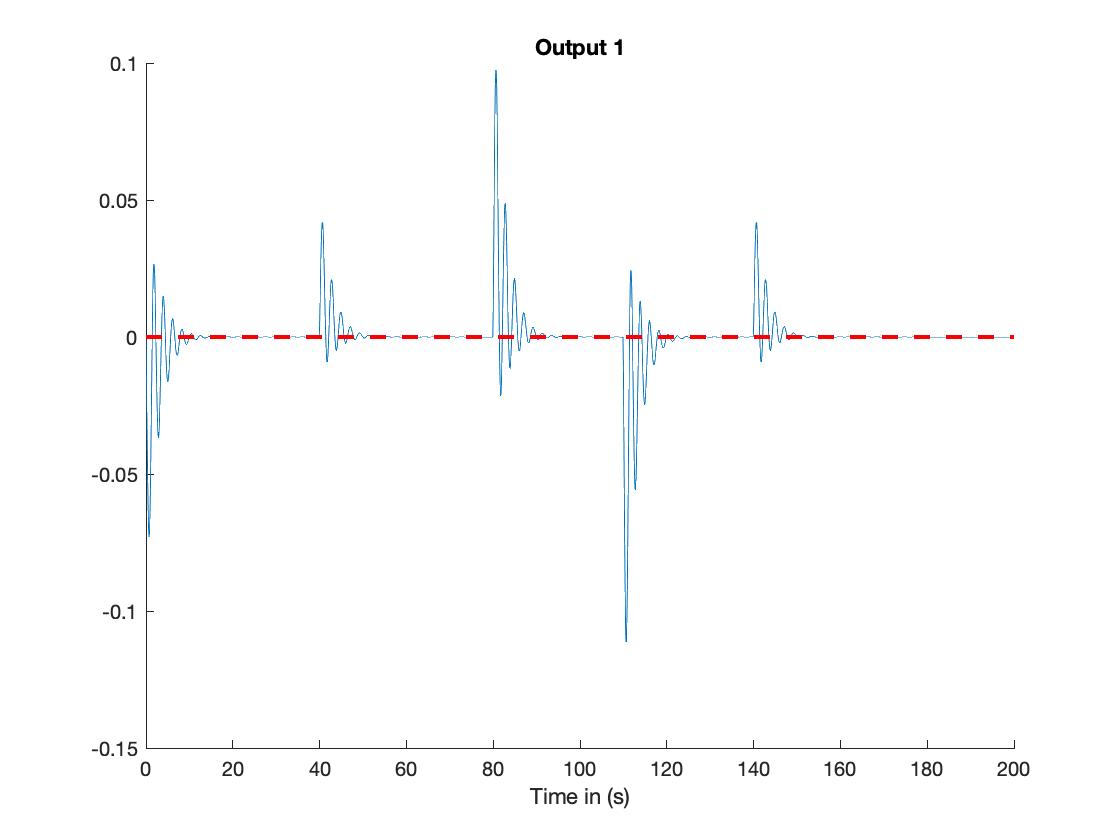,width=.29\textwidth}}
\centering%
\subfigure[\footnotesize Control line 1 ]
{\epsfig{figure=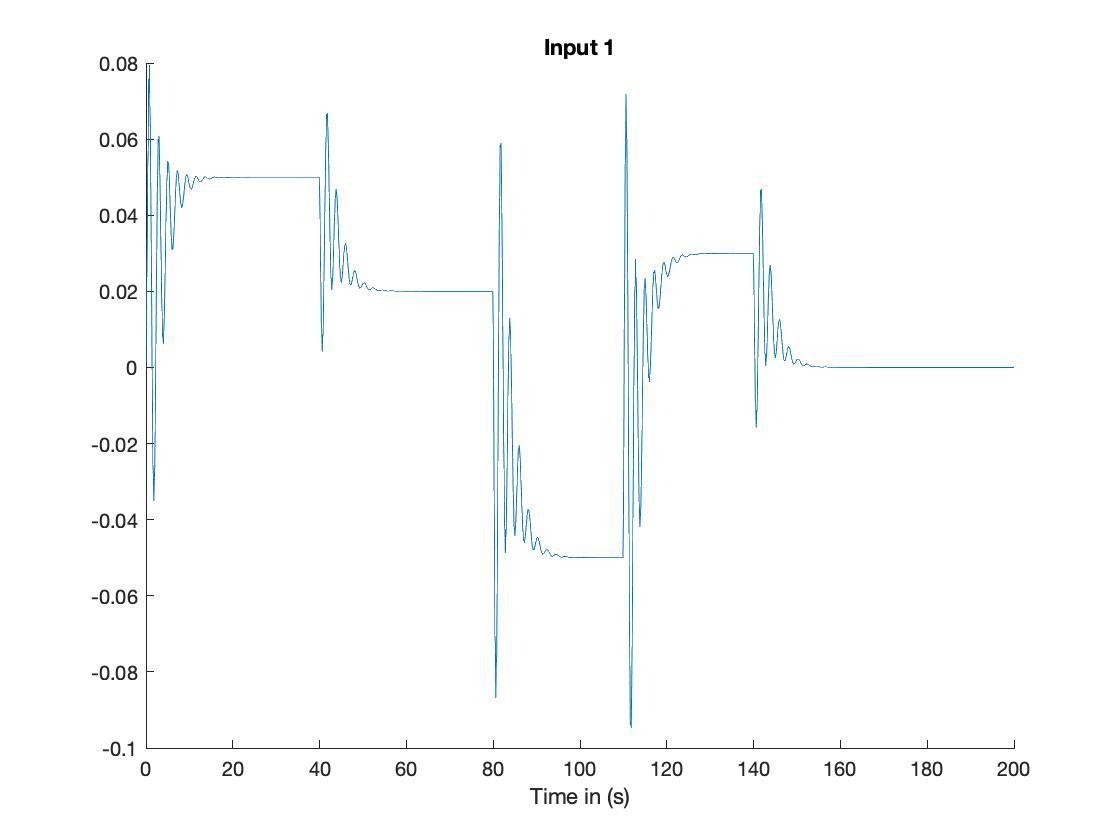,width=.29\textwidth}}
\\
\subfigure[\footnotesize Output (--) and reference trajectory (- -) line 2 ]
{\epsfig{figure=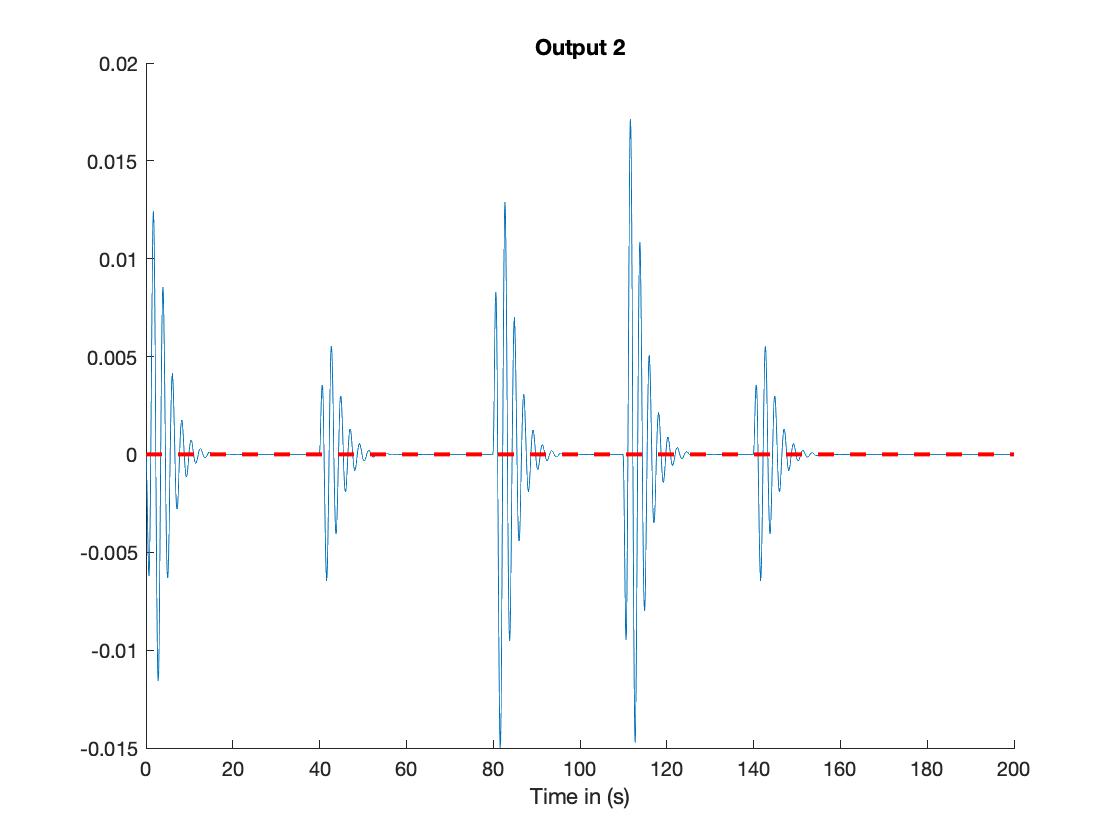,width=.29\textwidth}}
\centering%
\subfigure[\footnotesize Control line 2 ]
{\epsfig{figure=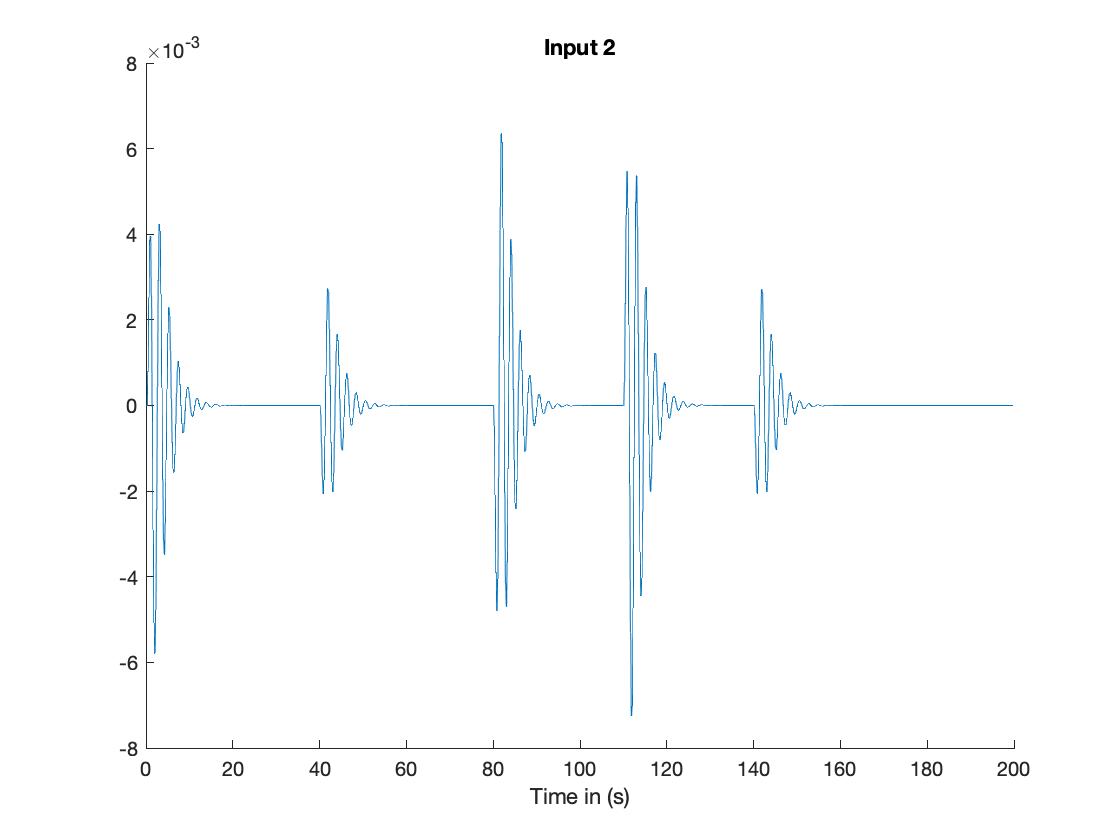,width=.29\textwidth}}
\caption{MFC: No attack}\label{CSM1}
\end{figure*}

Figures \ref{PID1} and \ref{CSM1}.
\subsubsection{Type 1 load altering attack}  Figures \ref{PID2} and \ref{CSM2}.
\subsubsection{Type 2 DoS attack with $90\%$ losses} Figures \ref{PID3} and \ref{CSM3}.
\subsubsection{Type 3 DoS attack with $95\%$ losses} 
\begin{figure*}[!ht]
\centering%
\subfigure[\footnotesize Output (--) and reference trajectory (- -) line 1 ]
{\epsfig{figure=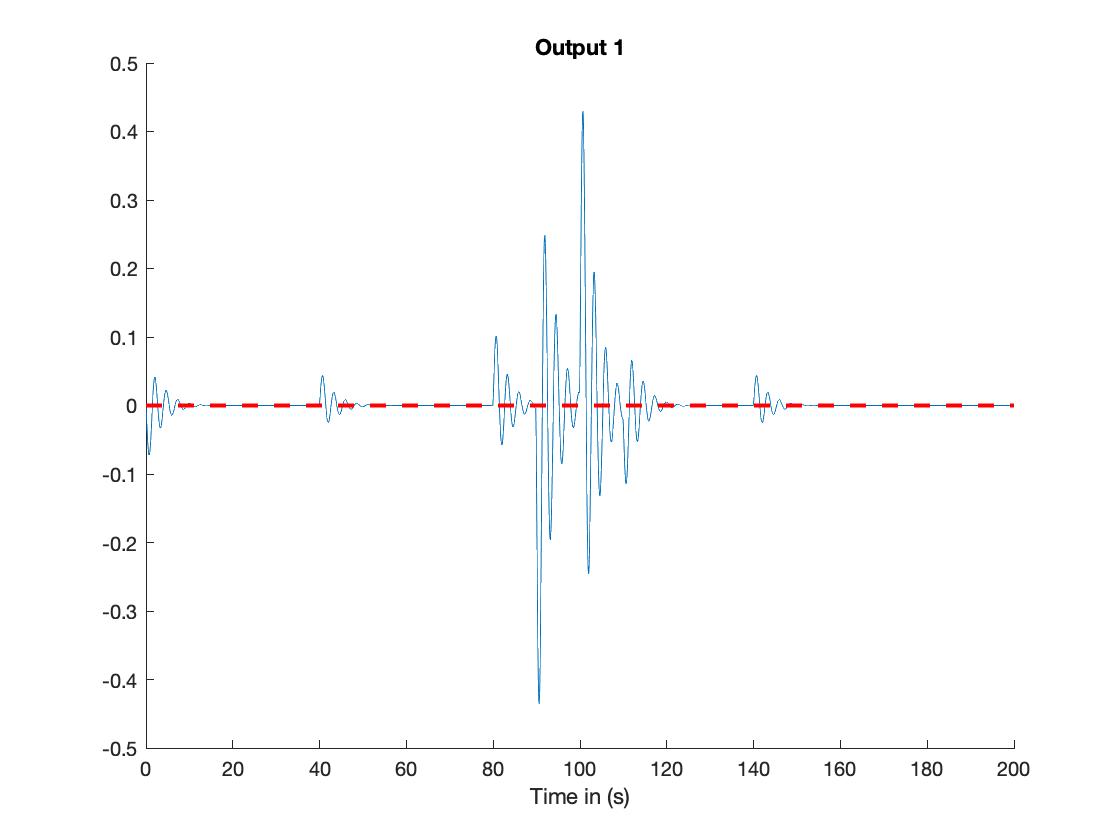,width=.29\textwidth}}
\centering%
\subfigure[\footnotesize Control line 1 ]
{\epsfig{figure=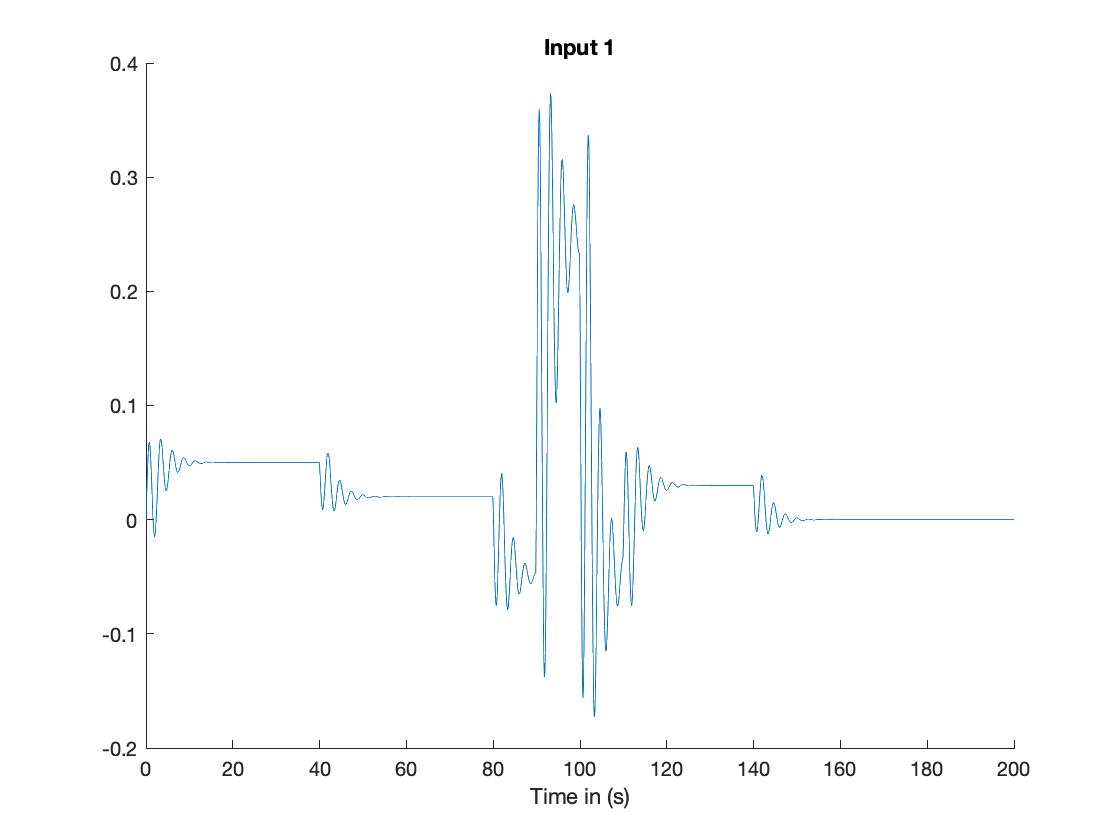,width=.29\textwidth}}
\\
\subfigure[\footnotesize Output (--) and reference trajectory (- -) line 2 ]
{\epsfig{figure=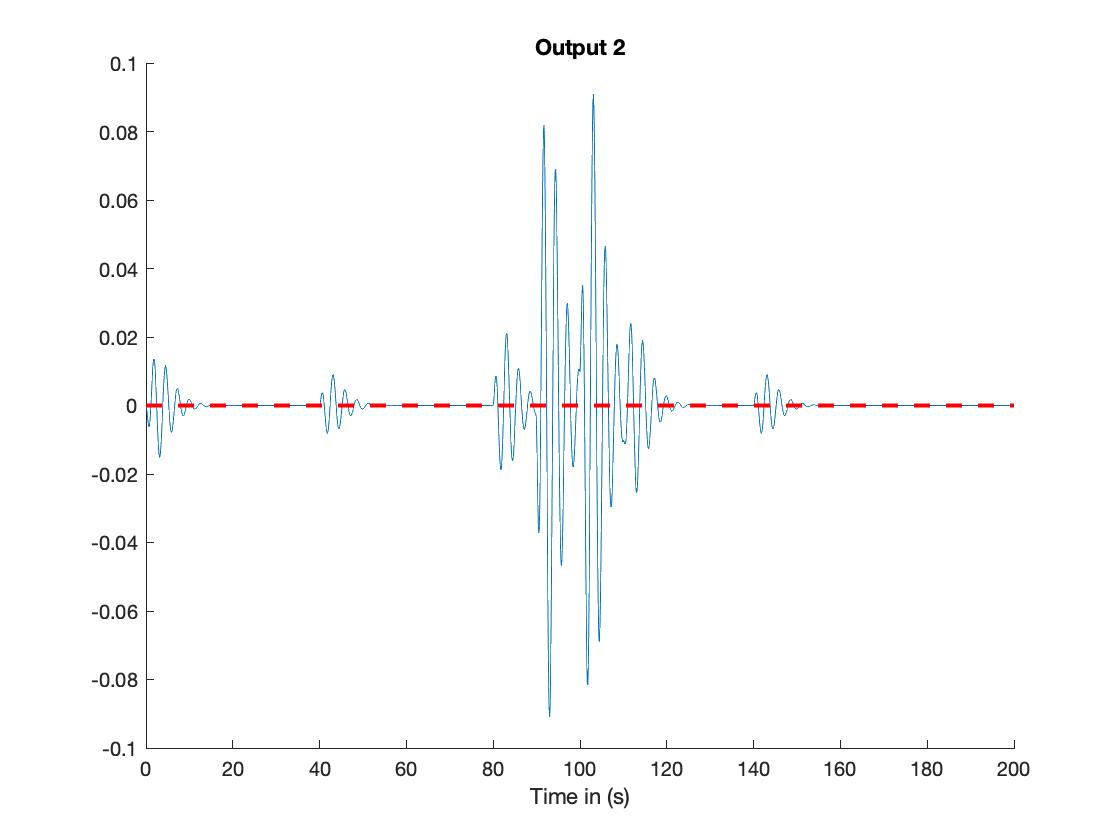,width=.29\textwidth}}
\centering%
\subfigure[\footnotesize Control line 2 ]
{\epsfig{figure=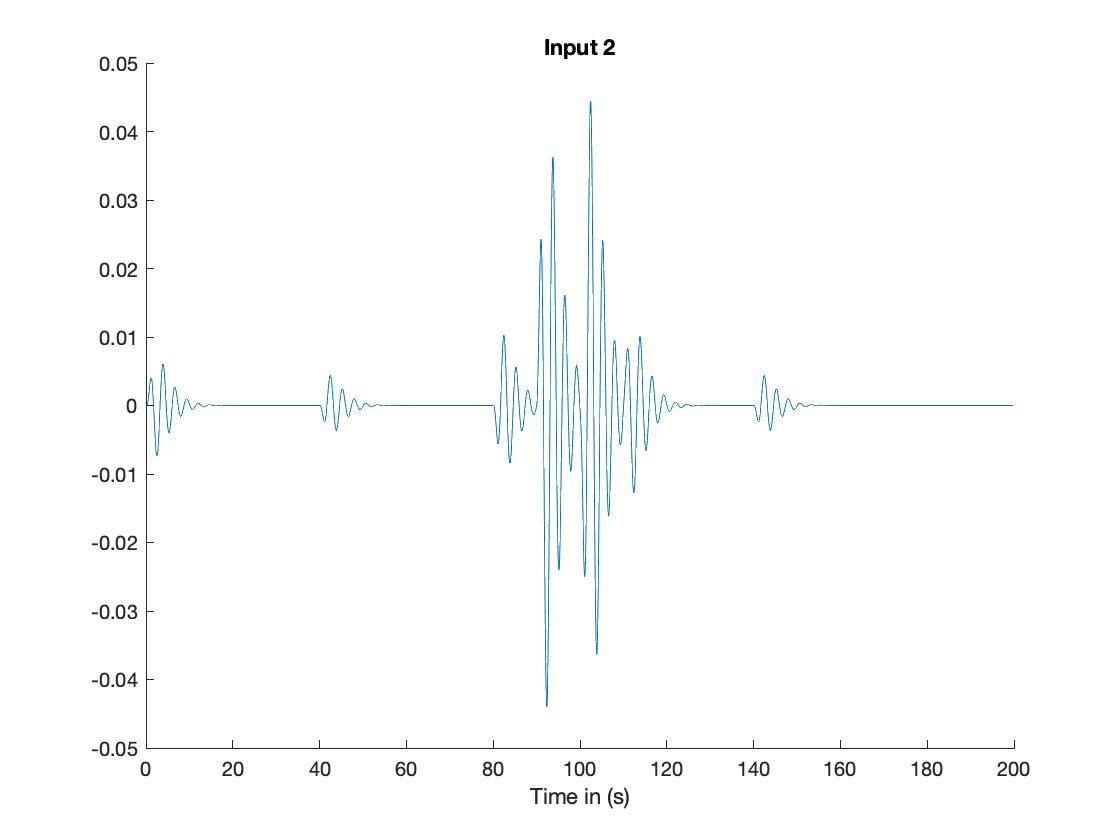,width=.29\textwidth}}
\subfigure[\footnotesize Abrupt change of power load]
{\epsfig{figure=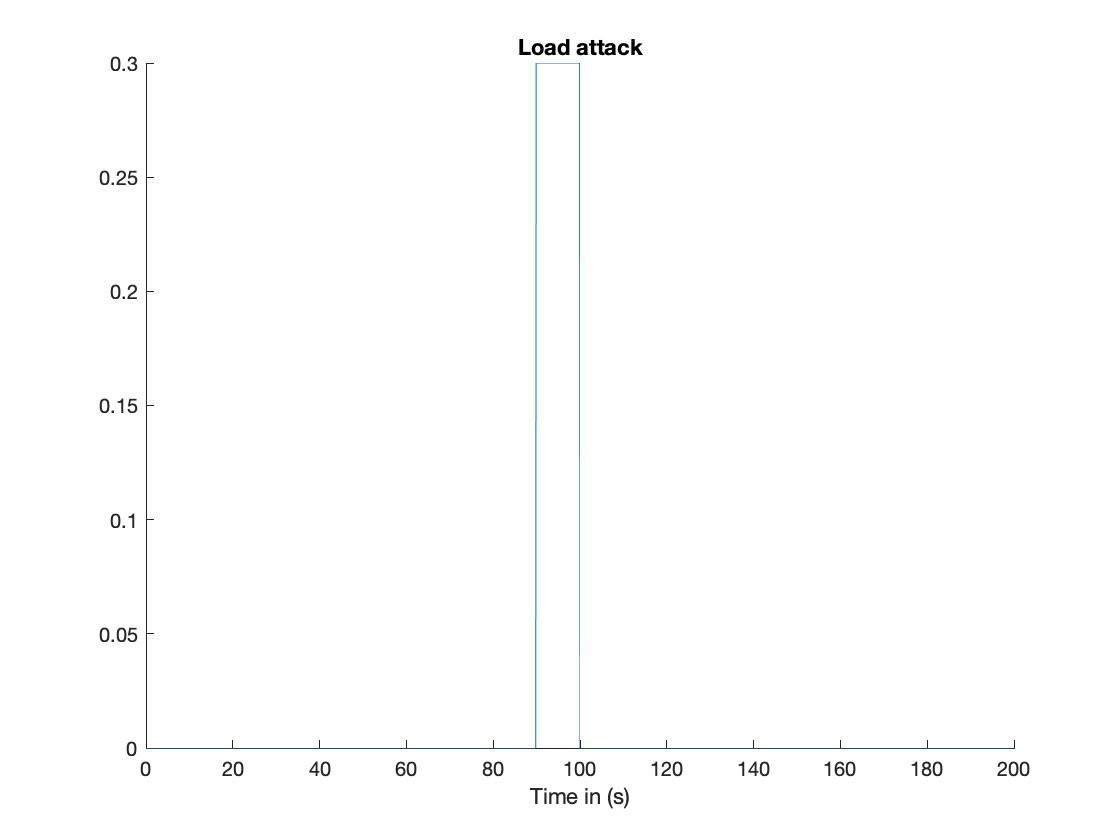,width=.29\textwidth}}
\caption{Integrator: Load altering attack}\label{PID2}
\end{figure*}

\begin{figure*}[!ht]
\centering%
\subfigure[\footnotesize Output (--) and reference trajectory (- -) line 1 ]
{\epsfig{figure=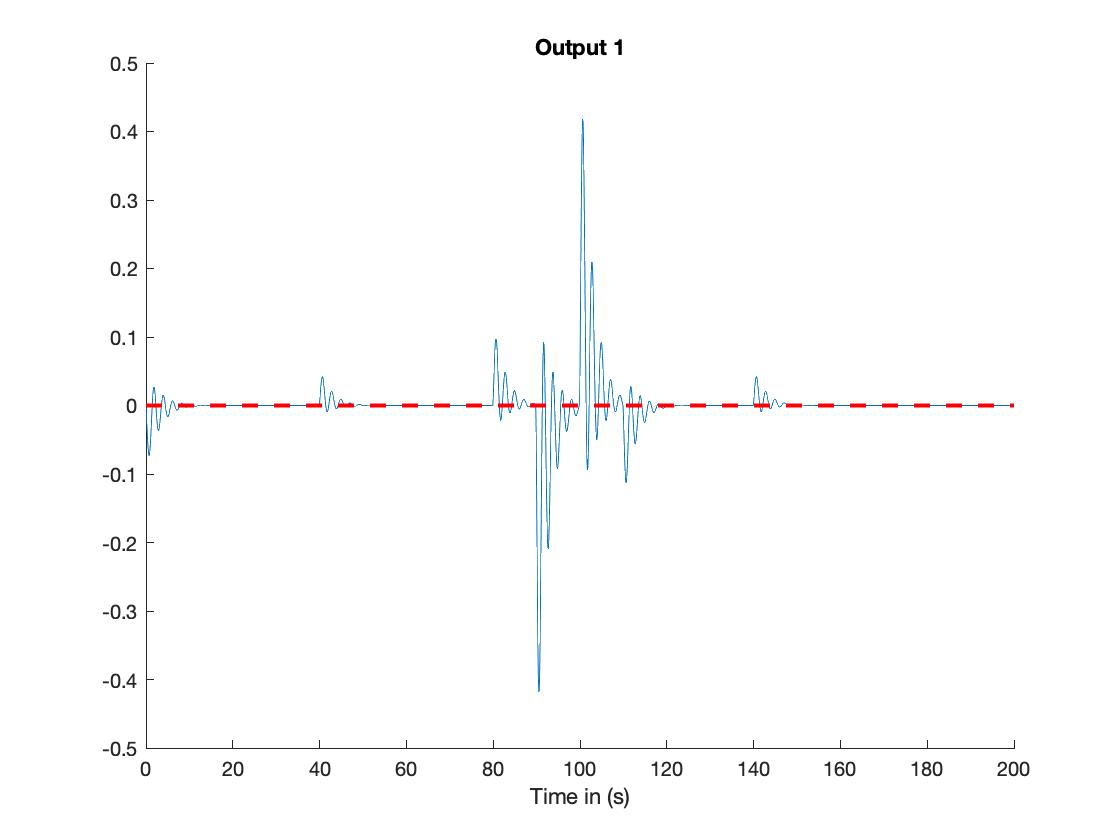,width=.29\textwidth}}
\centering%
\subfigure[\footnotesize Control line 1 ]
{\epsfig{figure=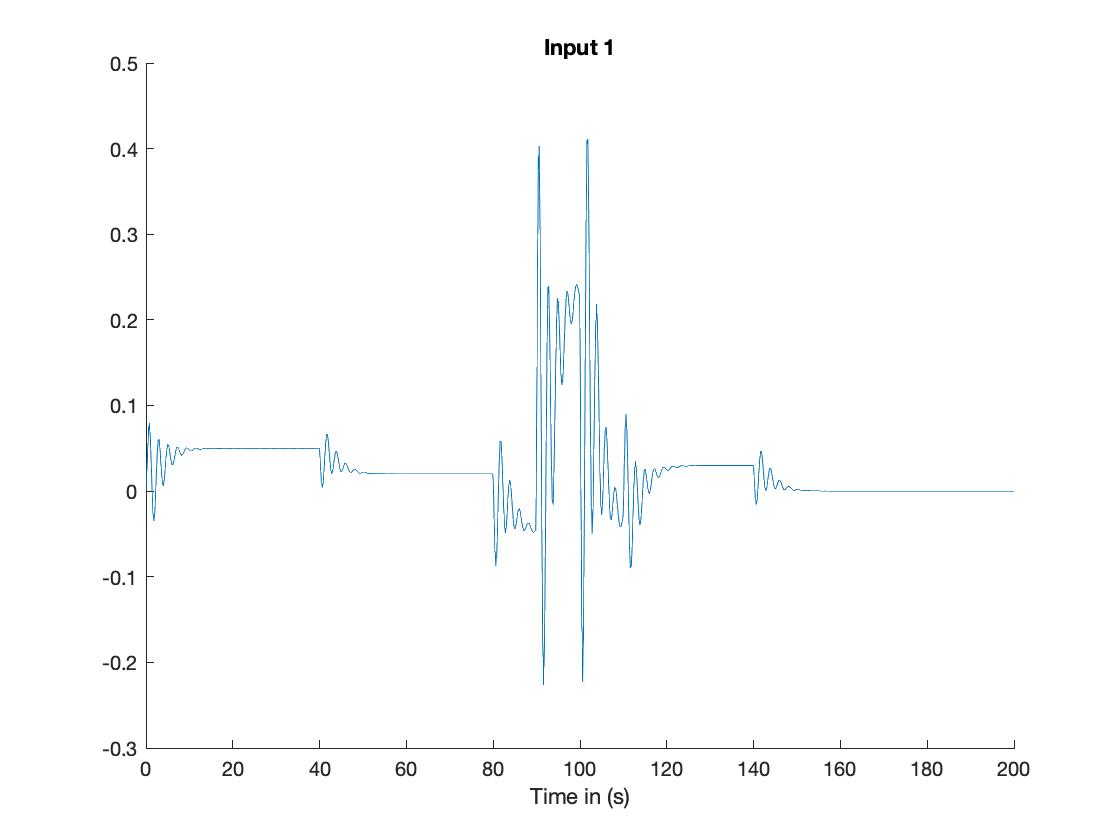,width=.29\textwidth}}
\\
\subfigure[\footnotesize Output (--) and reference trajectory (- -) line 2 ]
{\epsfig{figure=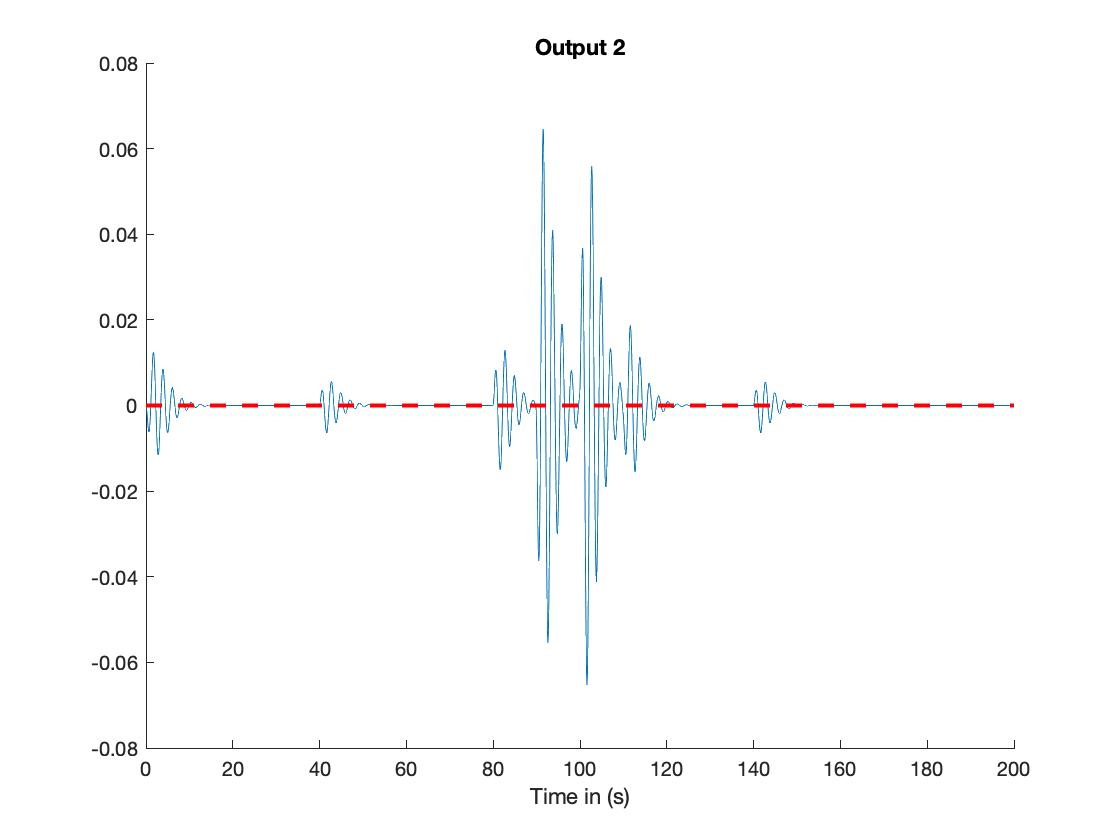,width=.29\textwidth}}
\centering%
\subfigure[\footnotesize Control line 2 ]
{\epsfig{figure=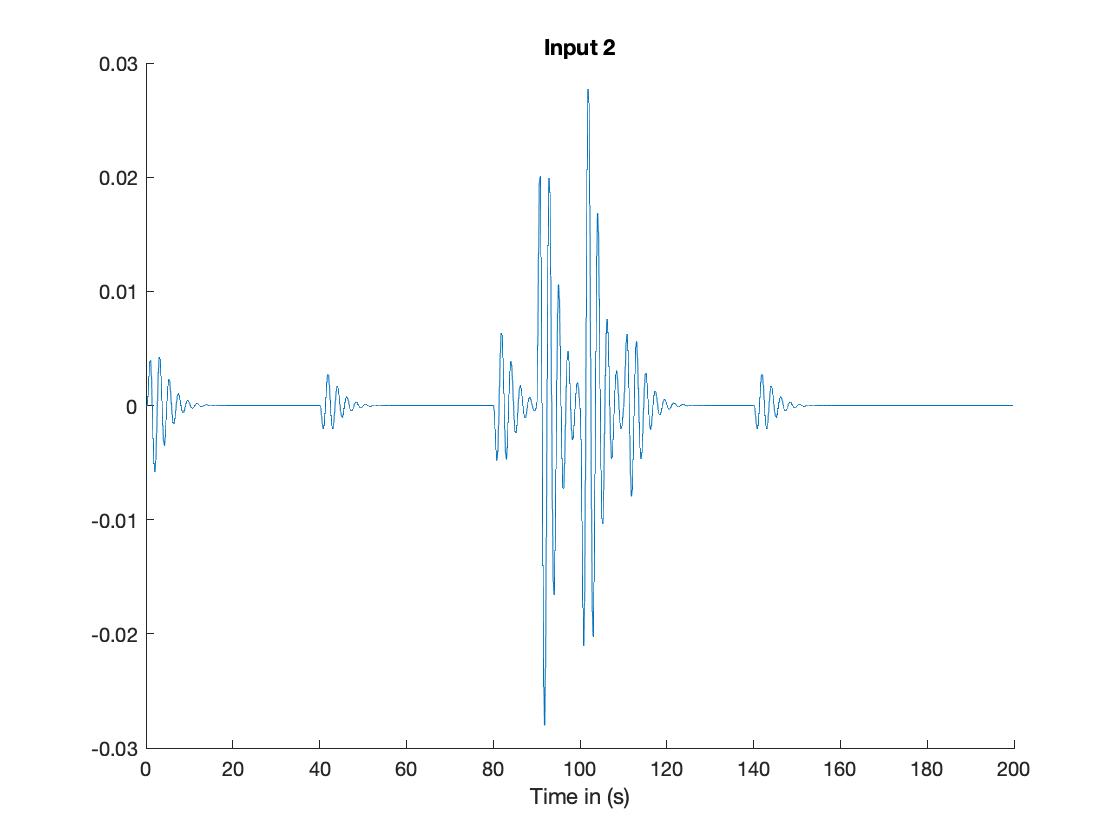,width=.29\textwidth}}
\subfigure[\footnotesize Abrupt change of power load]
{\epsfig{figure=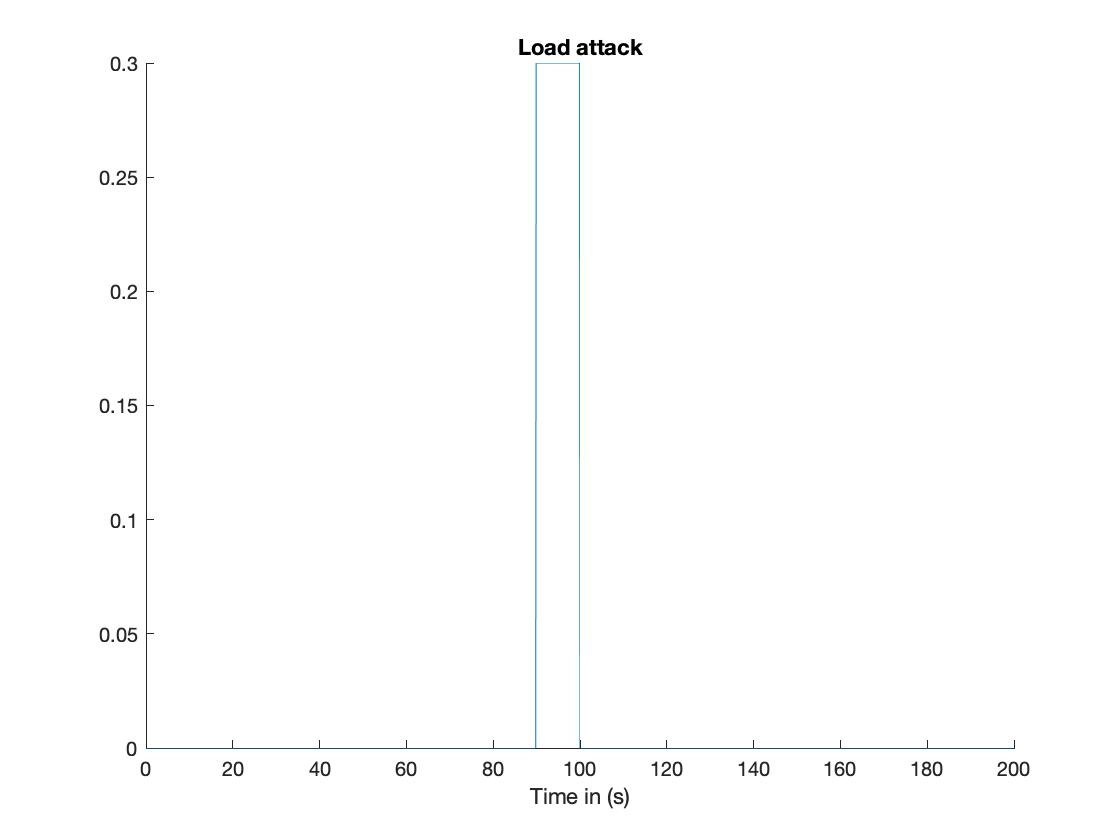,width=.29\textwidth}}
\caption{MFC: Load altering attack}\label{CSM2}
\end{figure*}

\begin{figure*}[!ht]
\centering%
\subfigure[\footnotesize Output (--) and reference trajectory (- -) line 1 ]
{\epsfig{figure=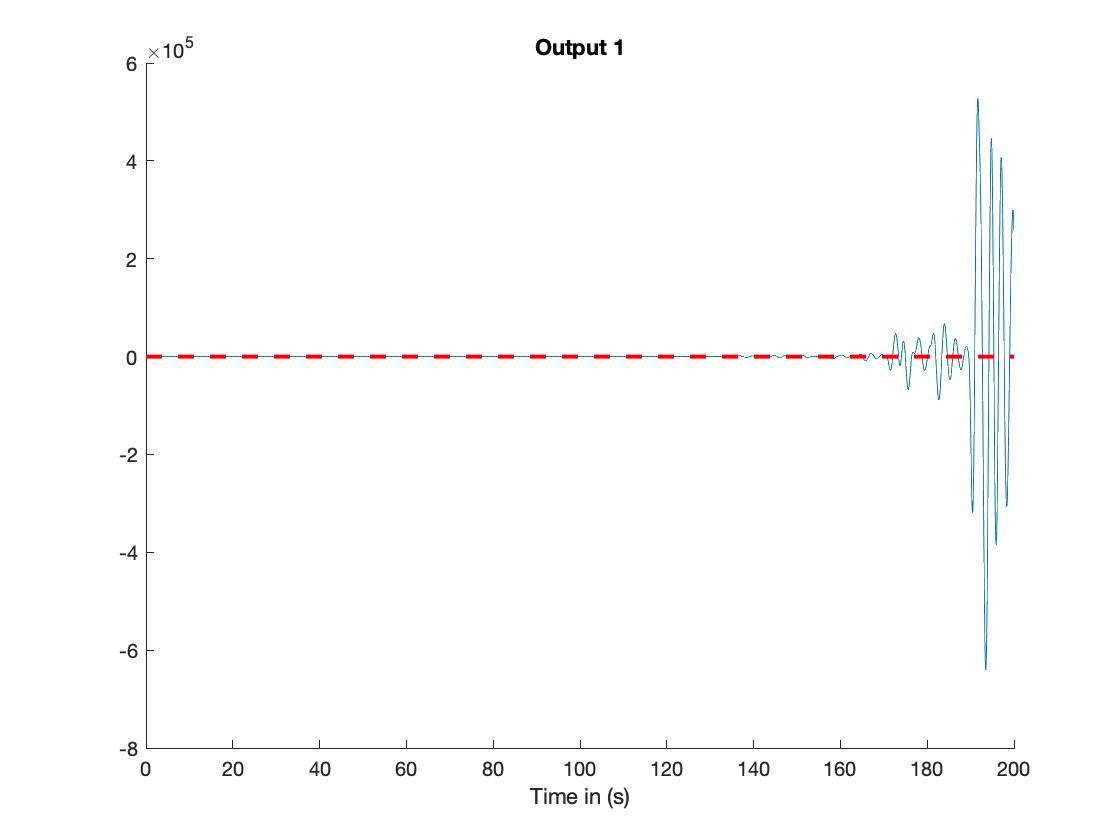,width=.29\textwidth}}
\centering%
\subfigure[\footnotesize Control line 1 ]
{\epsfig{figure=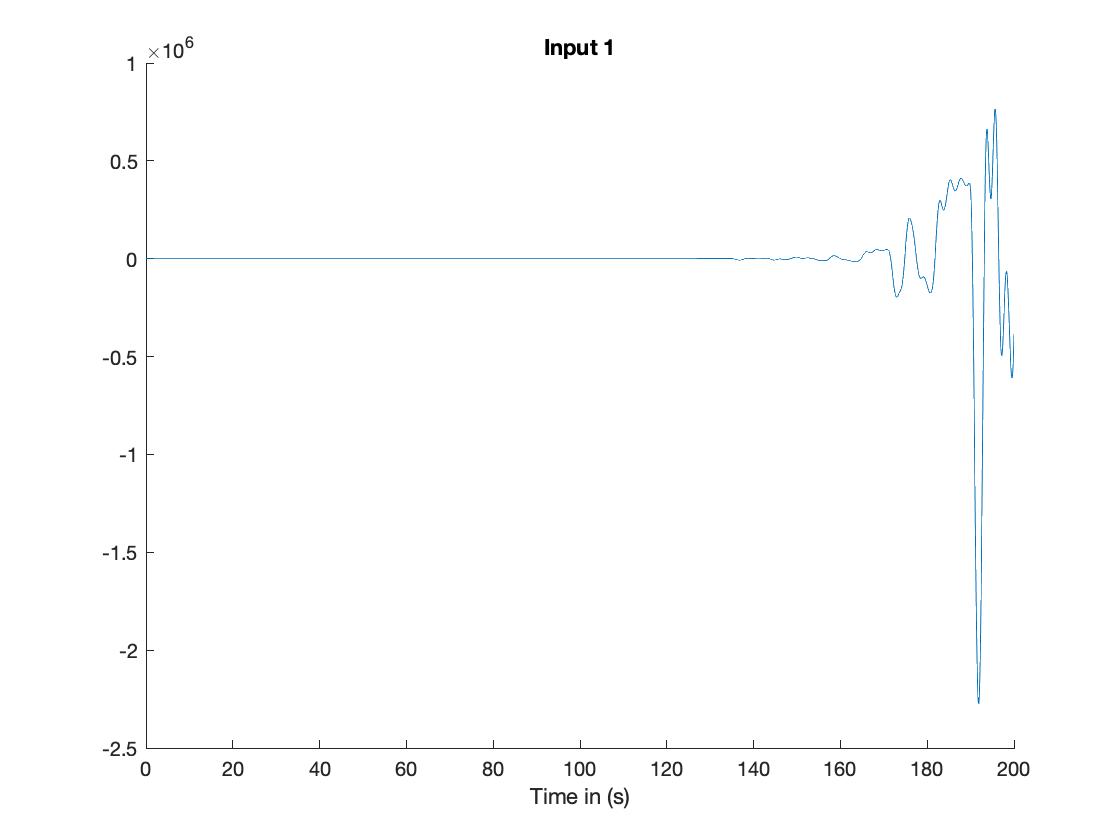,width=.29\textwidth}}
\subfigure[\footnotesize Output (--) and reference trajectory (- -) line 2 ]
{\epsfig{figure=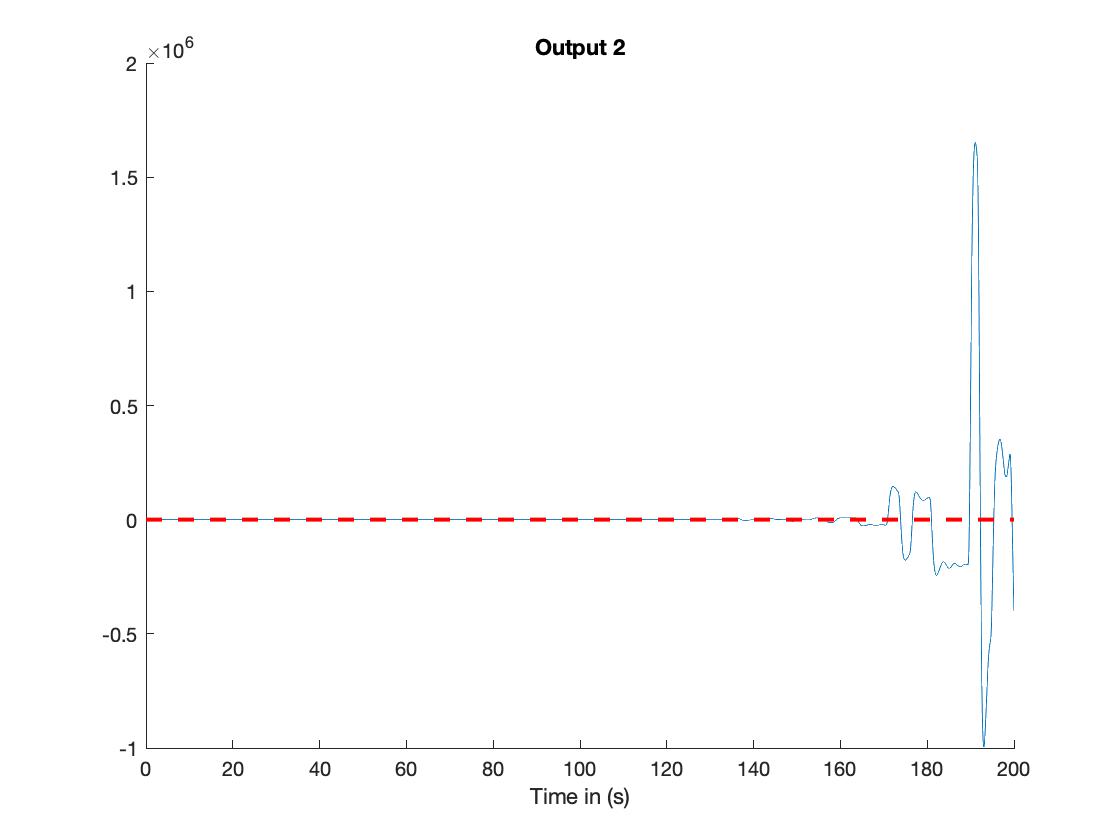,width=.29\textwidth}}
\centering%
\subfigure[\footnotesize Control line 2 ]
{\epsfig{figure=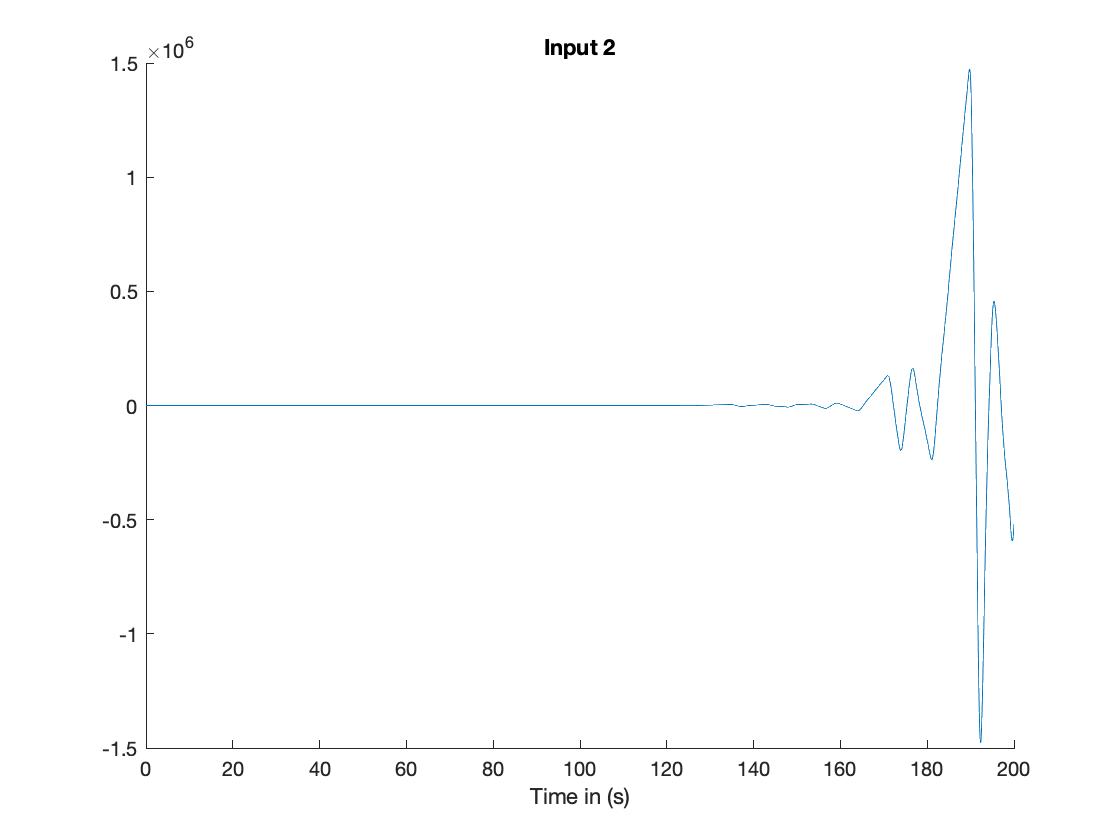,width=.29\textwidth}}
\subfigure[\footnotesize Attack-free control (red) and control after attack (blue)]
{\epsfig{figure=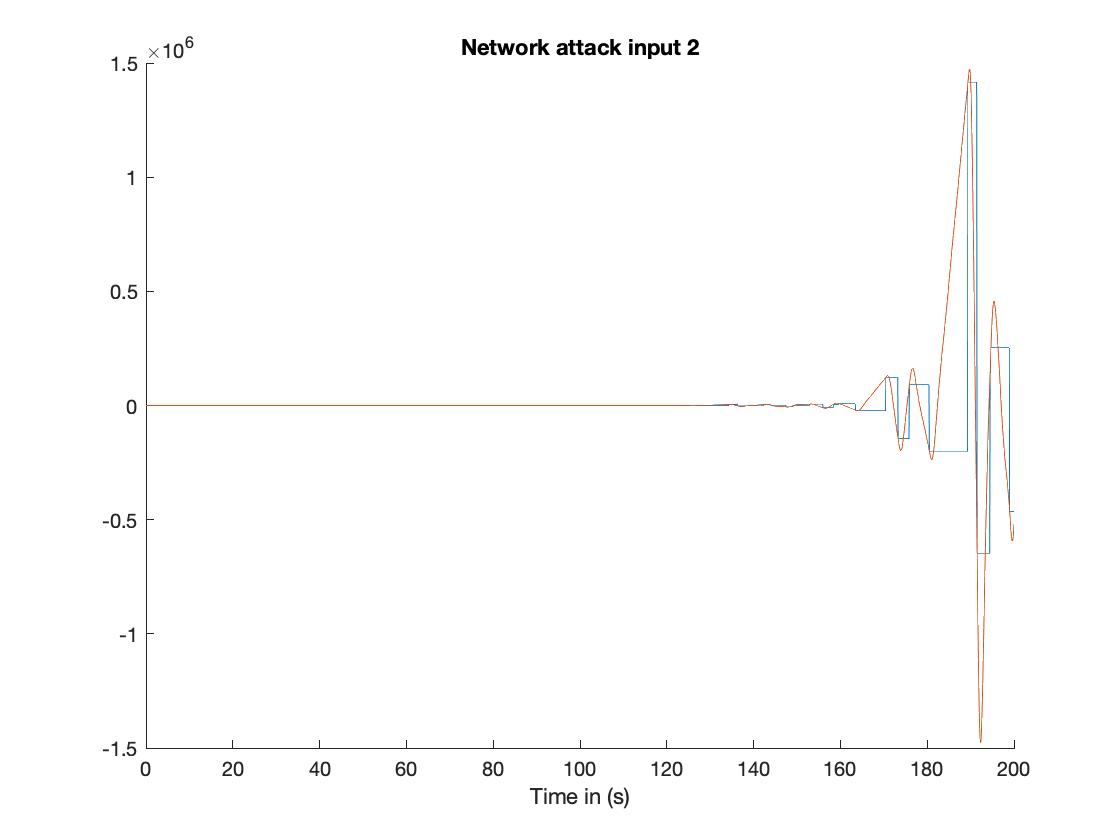,width=.29\textwidth}}
\subfigure[\footnotesize Zoom on (c)]
{\epsfig{figure=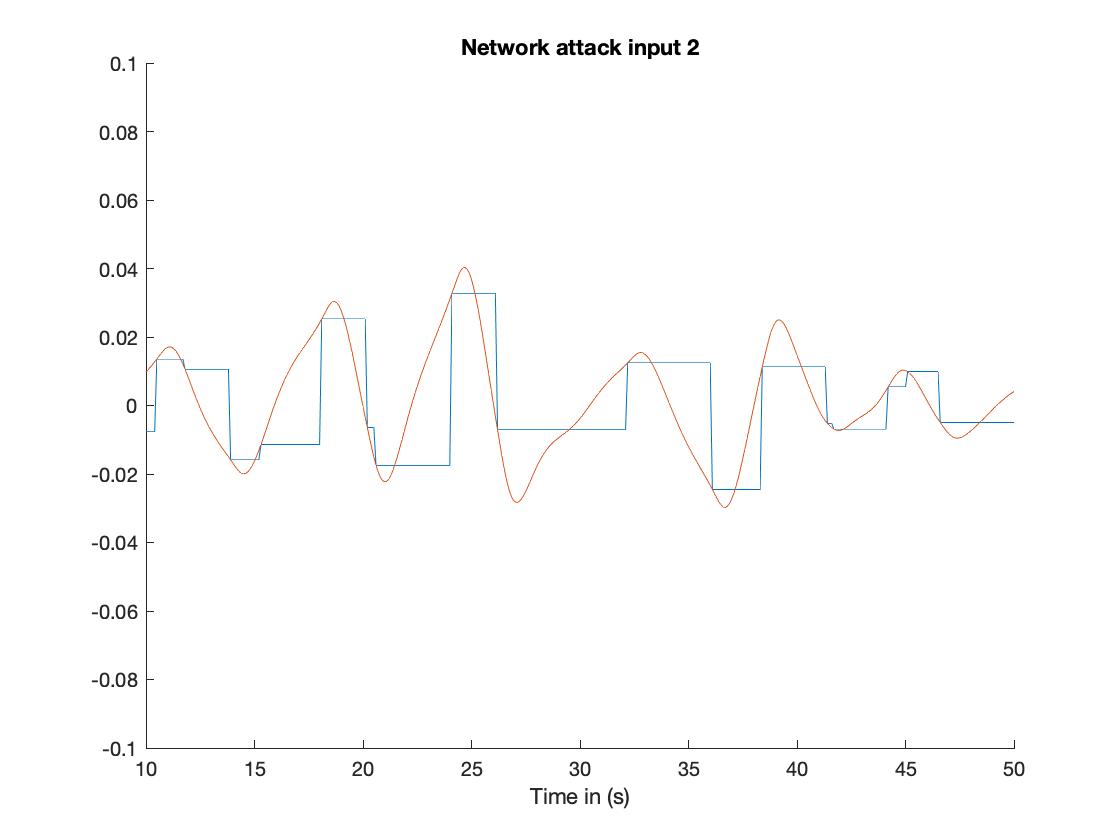,width=.29\textwidth}}
\caption{Integrator: Type 2 DoS attack}\label{PID3}
\end{figure*}

\begin{figure*}[!ht]
\centering%
\subfigure[\footnotesize Output (--) and reference trajectory (- -) line 1 ]
{\epsfig{figure=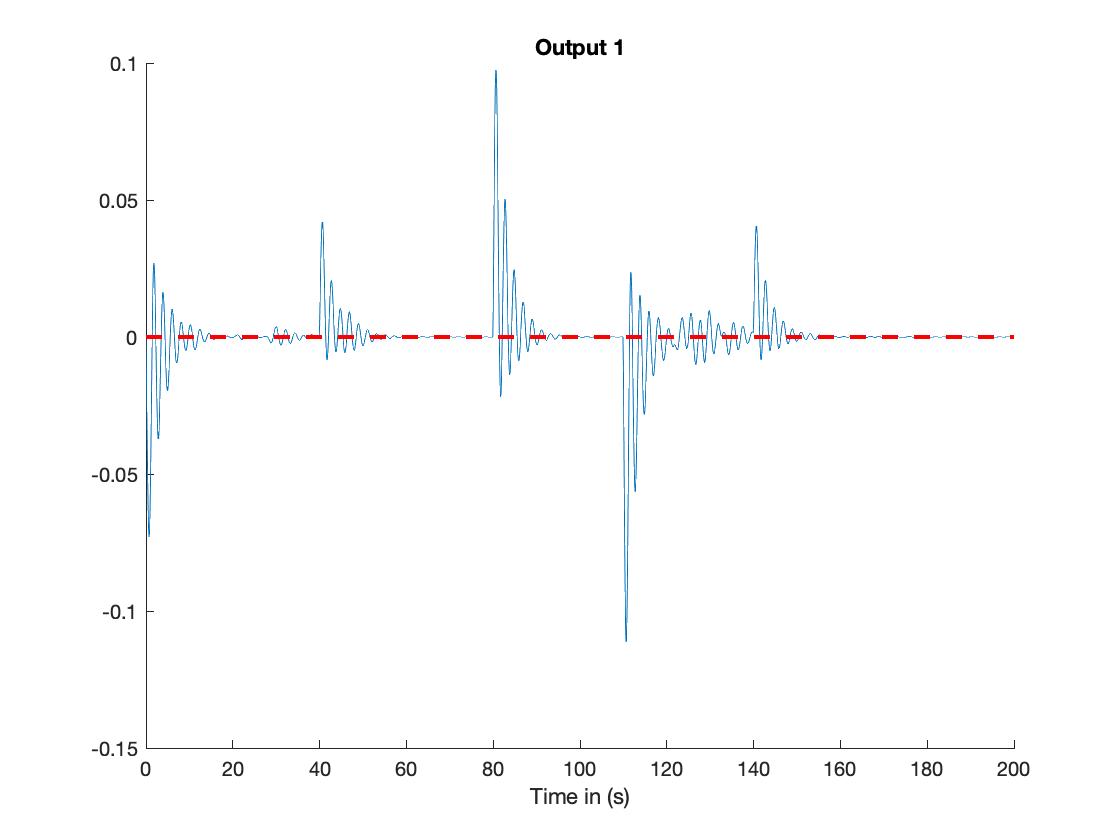,width=.29\textwidth}}
\centering%
\subfigure[\footnotesize Control line 1 ]
{\epsfig{figure=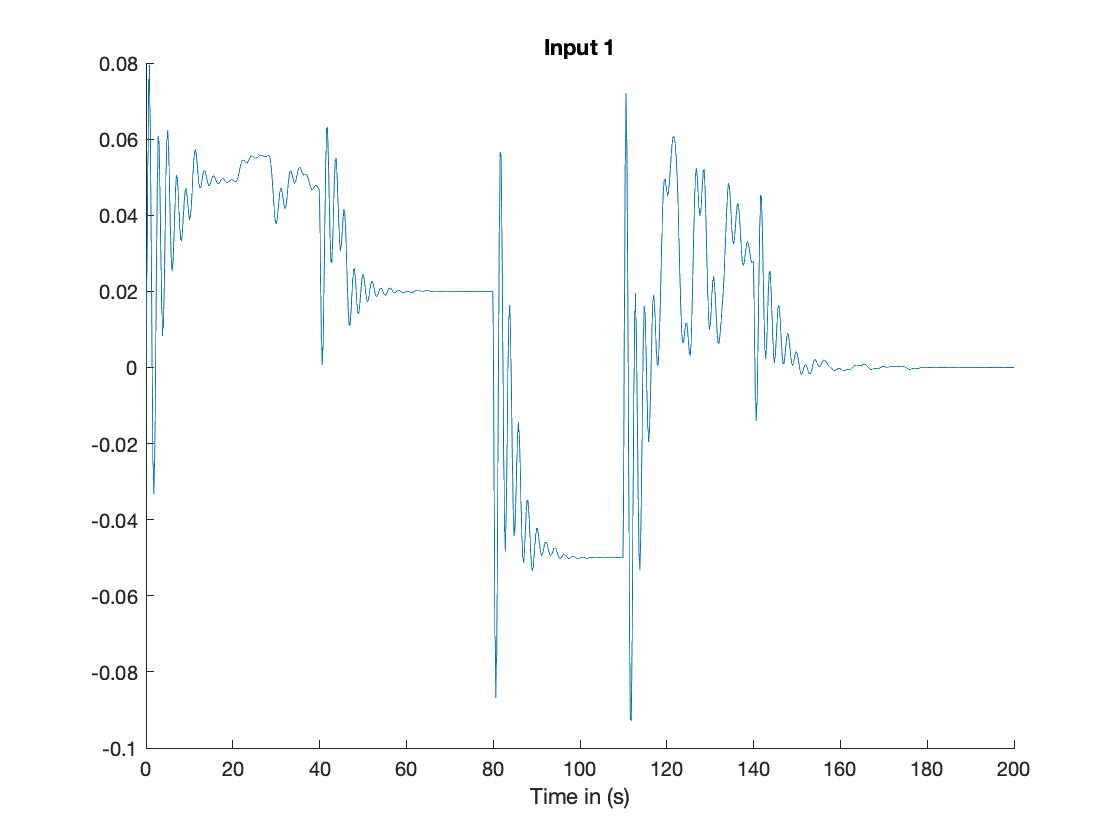,width=.29\textwidth}}
\subfigure[\footnotesize Output (--) and reference trajectory (- -) line 2 ]
{\epsfig{figure=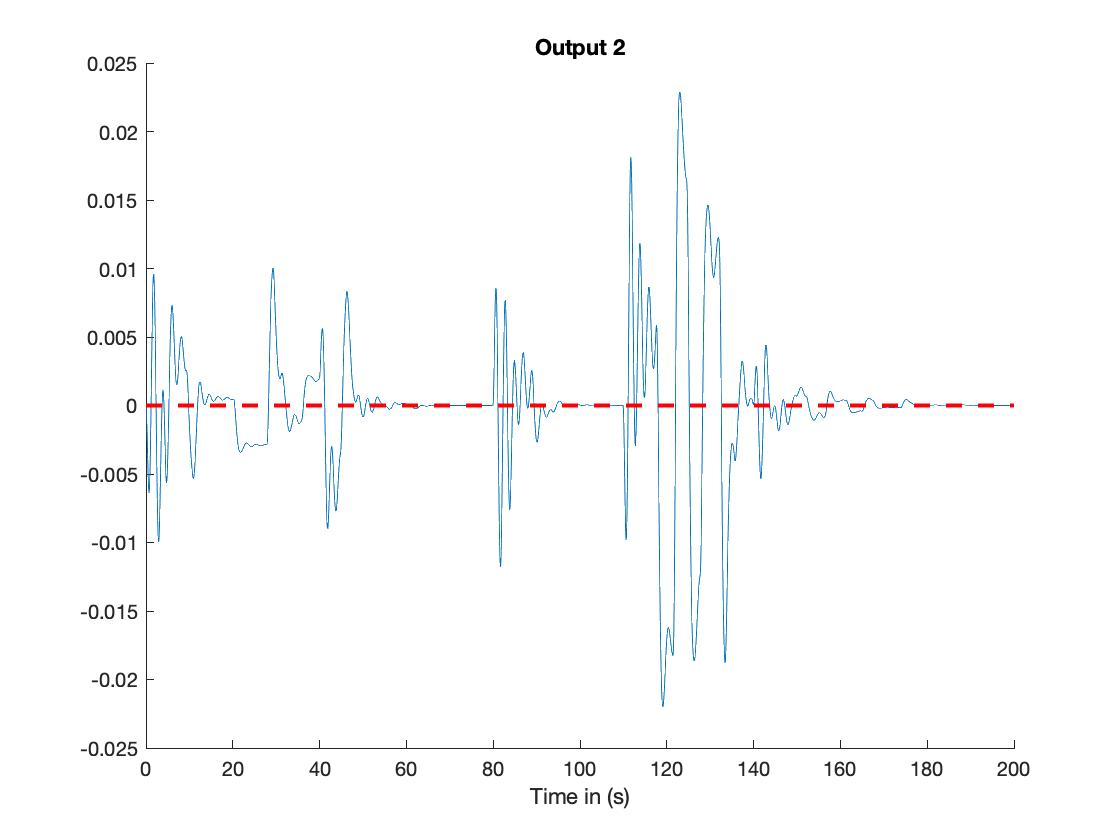,width=.29\textwidth}}
\centering%
\subfigure[\footnotesize Control line 2 ]
{\epsfig{figure=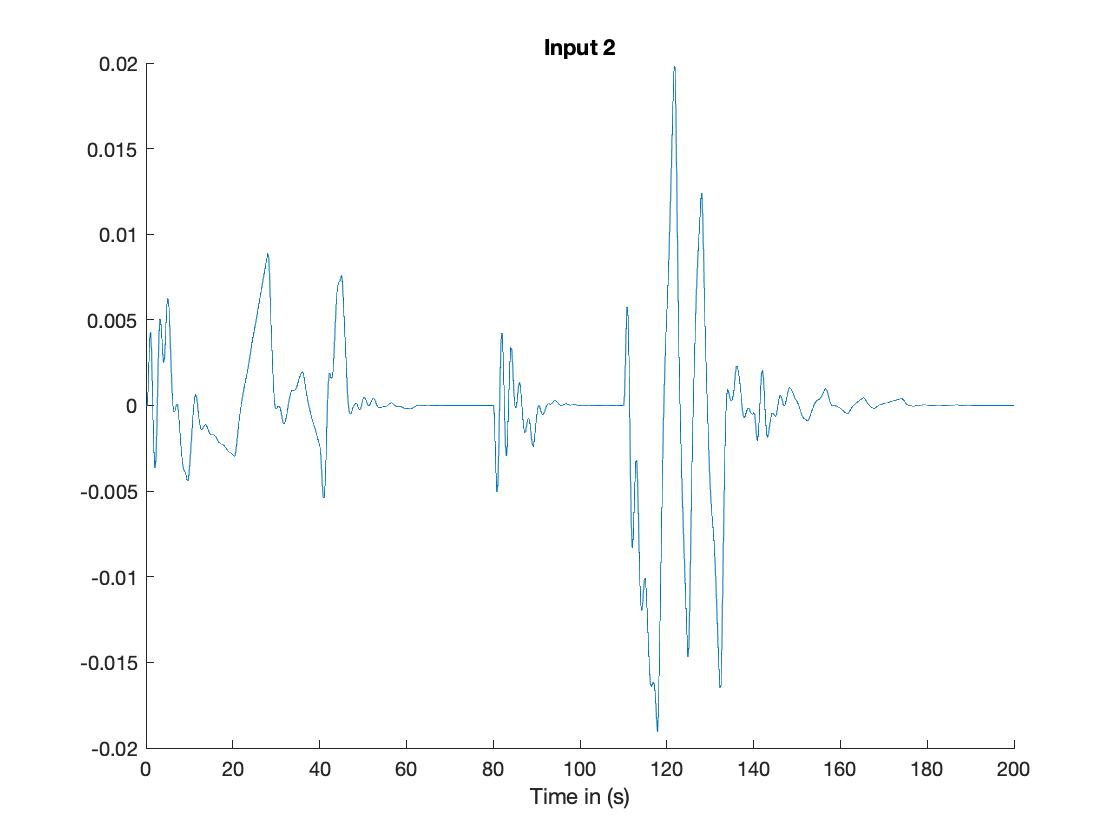,width=.29\textwidth}}
\subfigure[\footnotesize Attack-free control (red) and control after attack (blue)]
{\epsfig{figure=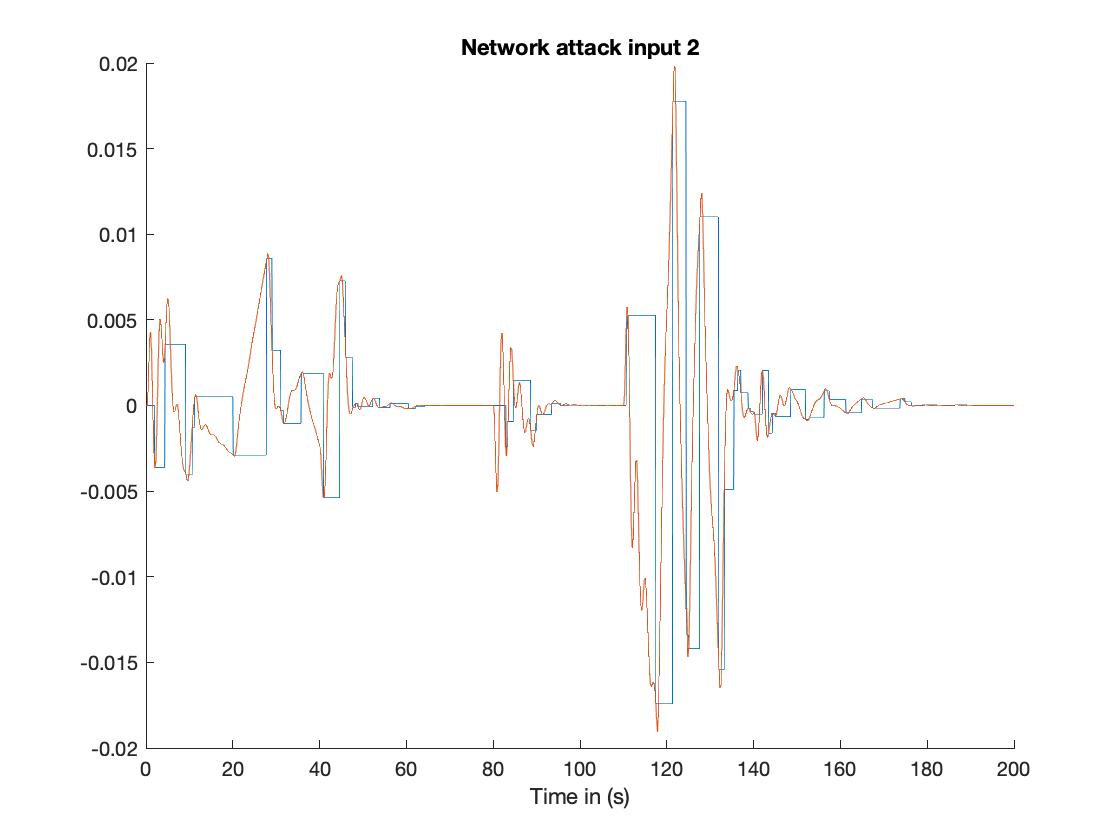,width=.29\textwidth}}
\subfigure[\footnotesize Zoom on (c)]
{\epsfig{figure=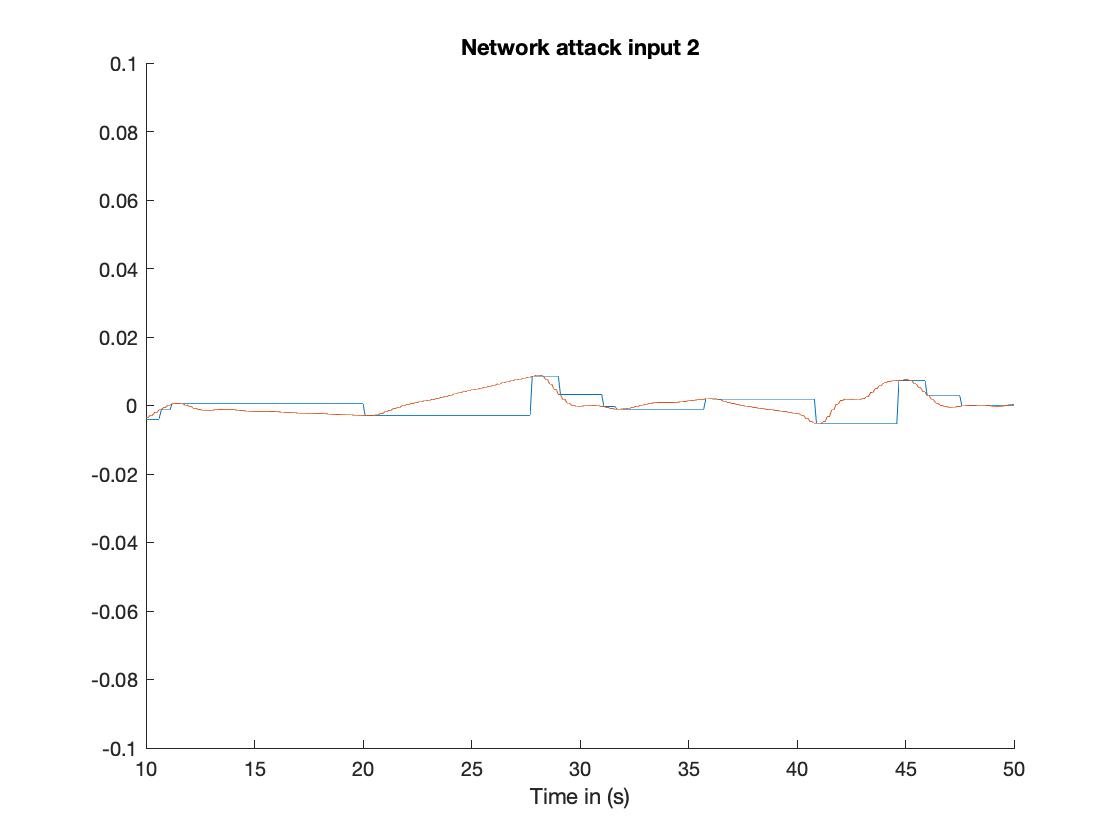,width=.29\textwidth}}
\caption{MFC: Type 2 DoS attack}\label{CSM3}
\end{figure*}

\begin{figure*}[!ht]
\centering%
\subfigure[\footnotesize Output (--) and reference trajectory (- -) line 1 ]
{\epsfig{figure=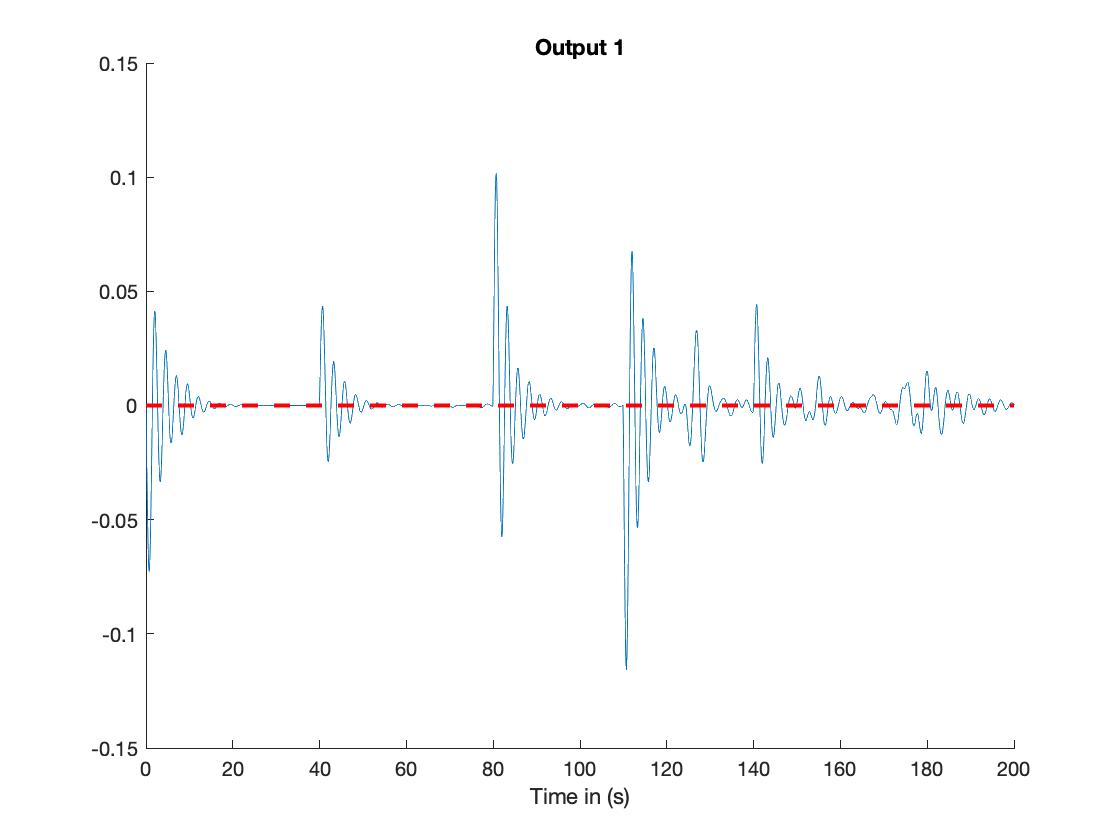,width=.29\textwidth}}
\centering%
\subfigure[\footnotesize Control line 1 ]
{\epsfig{figure=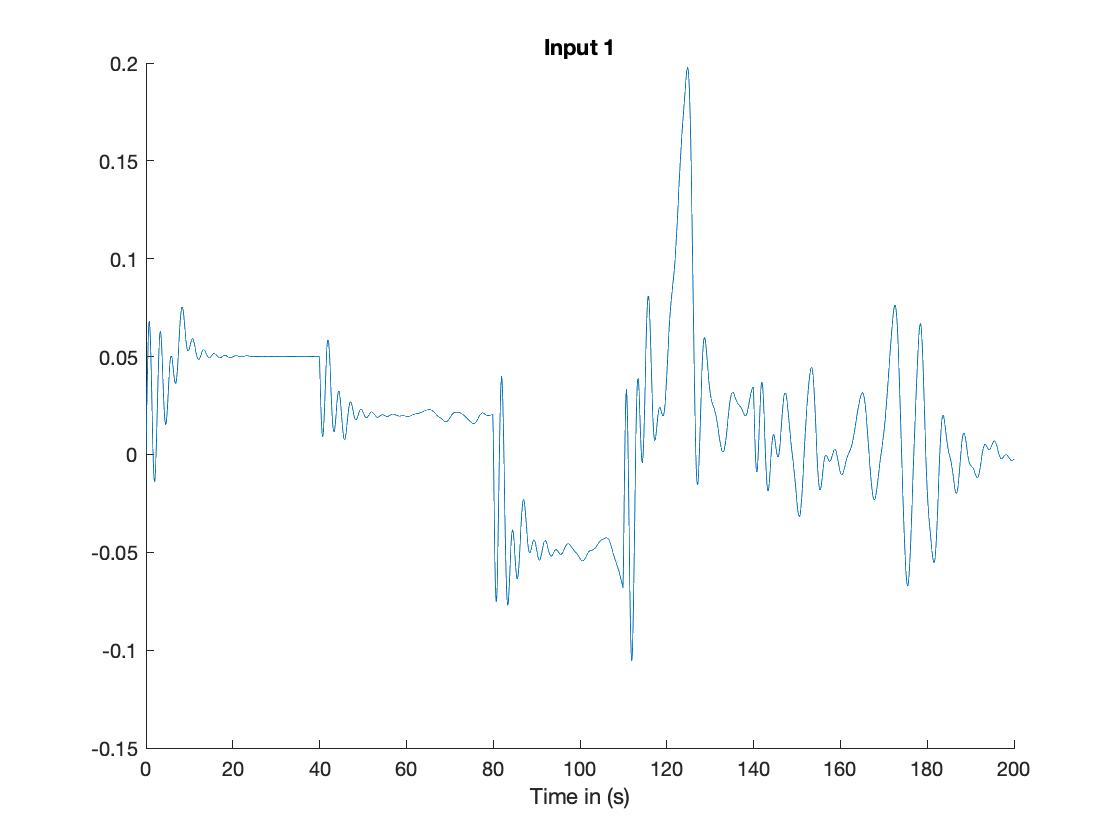,width=.29\textwidth}}
\subfigure[\footnotesize Output (--) and reference trajectory (- -) line 2 ]
{\epsfig{figure=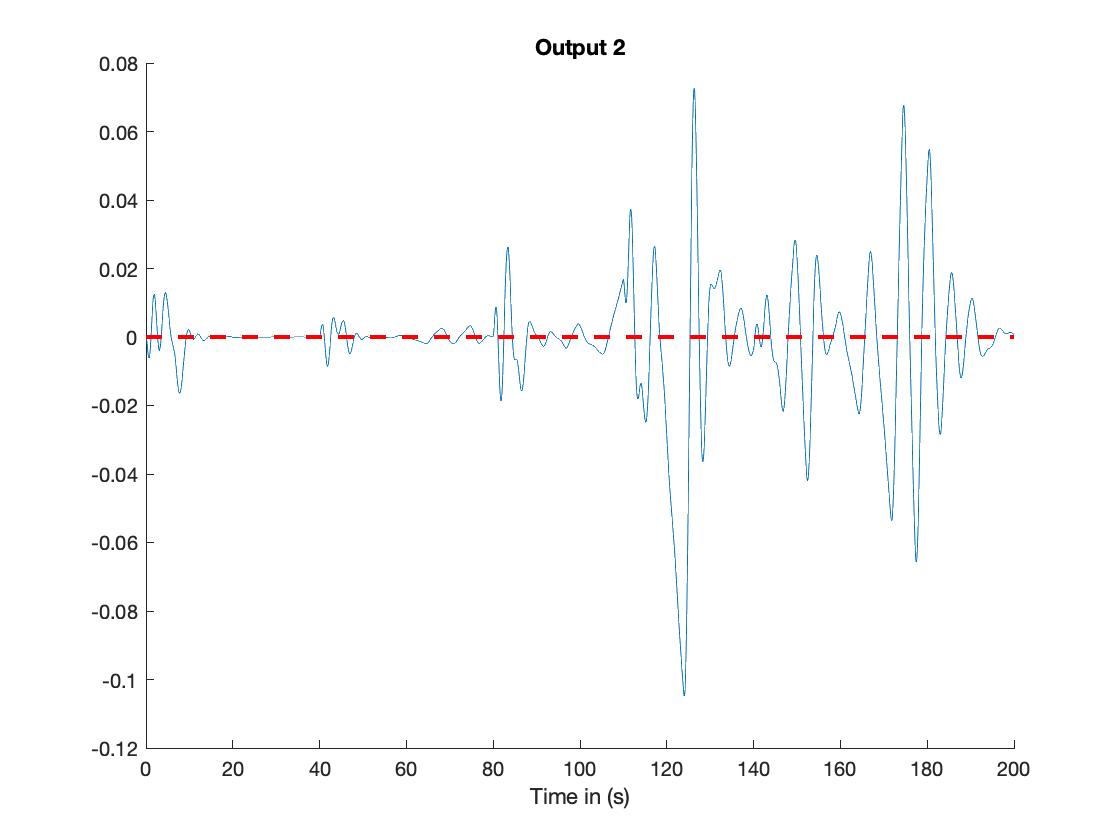,width=.29\textwidth}}
\centering%
\subfigure[\footnotesize Control line 2 ]
{\epsfig{figure=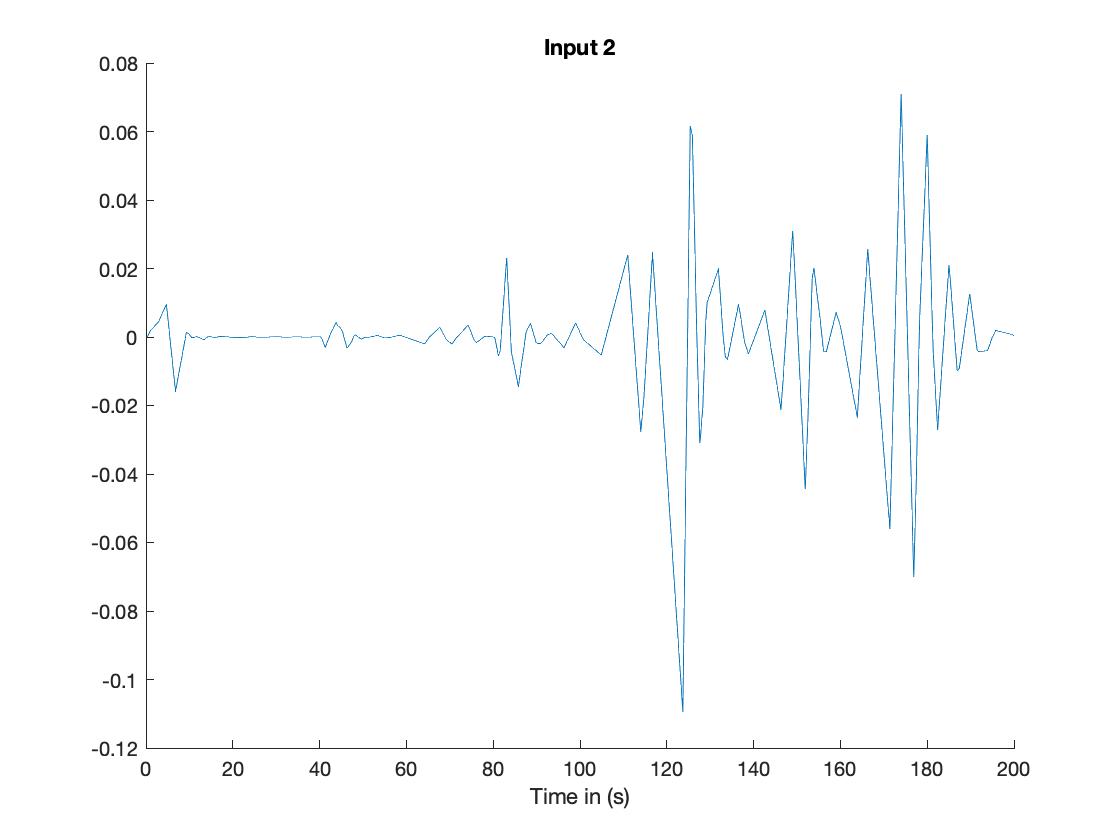,width=.29\textwidth}}
\subfigure[\footnotesize Attack-free measured output (red) and measured output after attack (blue)]
{\epsfig{figure=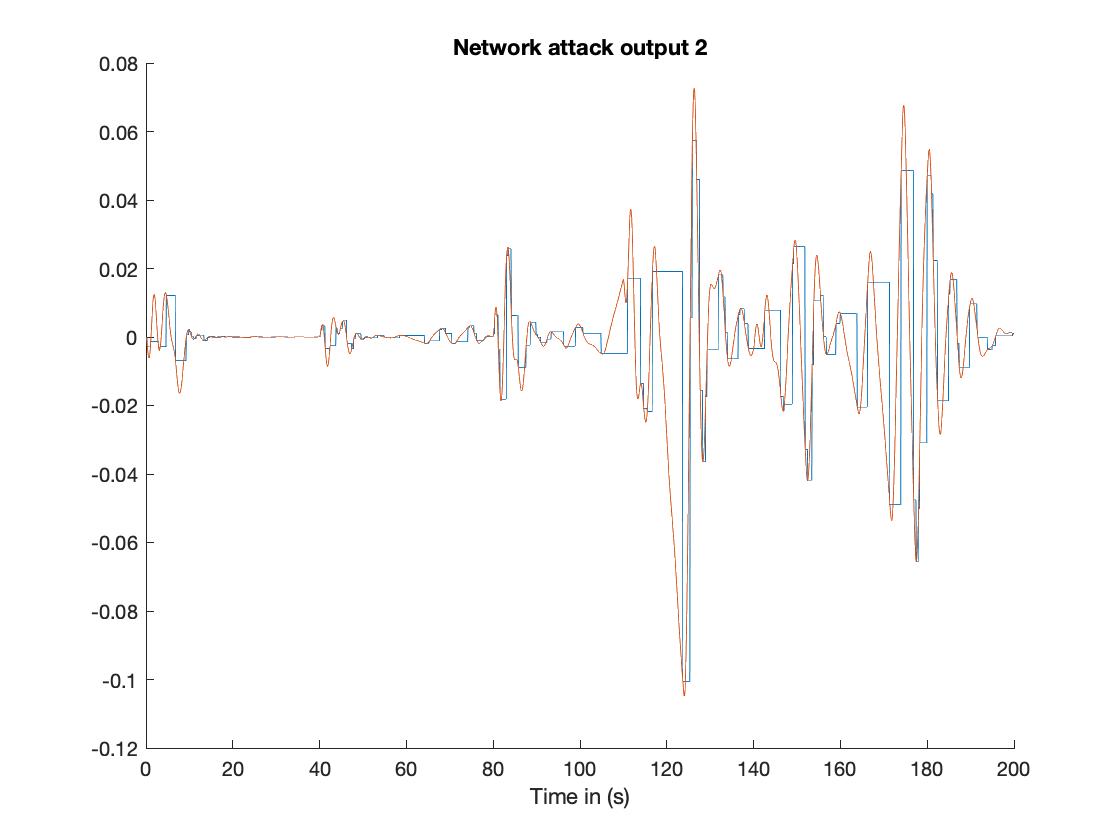,width=.29\textwidth}}
\subfigure[\footnotesize Zoom on (c)]
{\epsfig{figure=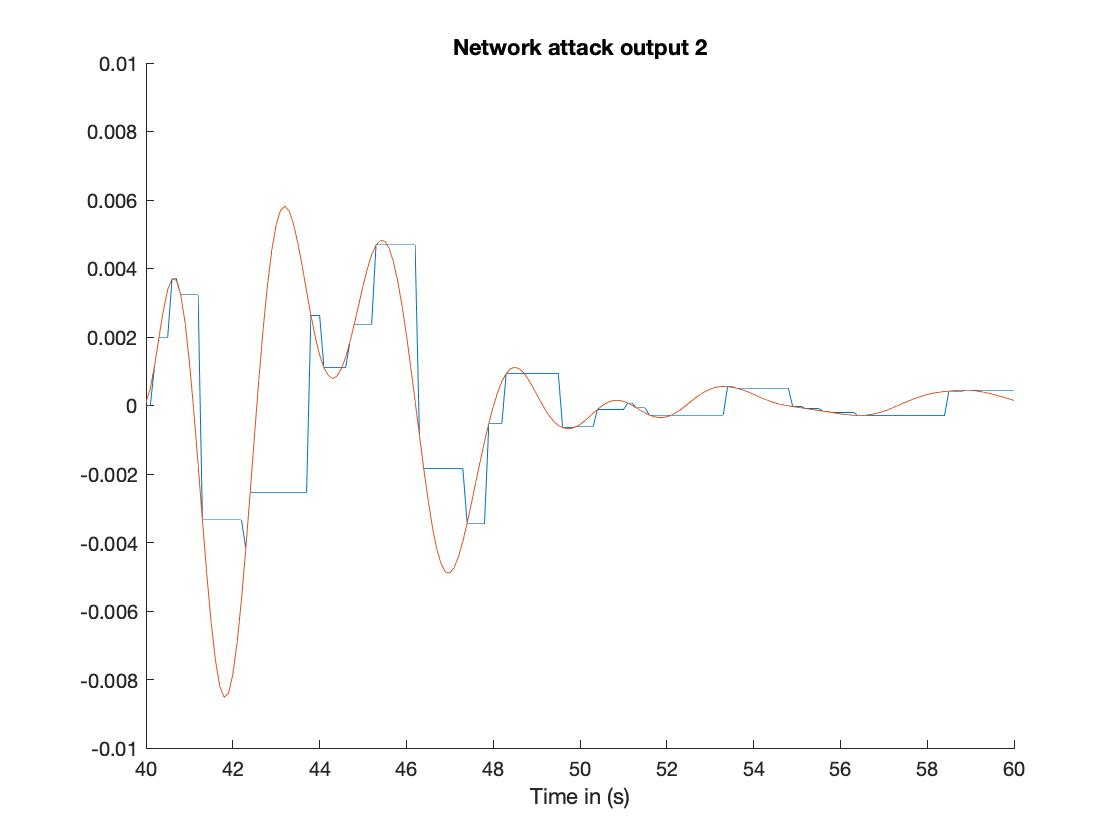,width=.29\textwidth}}
\caption{Integrator: Type 3 DoS attack}\label{PID4}
\end{figure*}

\begin{figure*}[!ht]
\centering%
\subfigure[\footnotesize Output (--) and reference trajectory (- -) line 1 ]
{\epsfig{figure=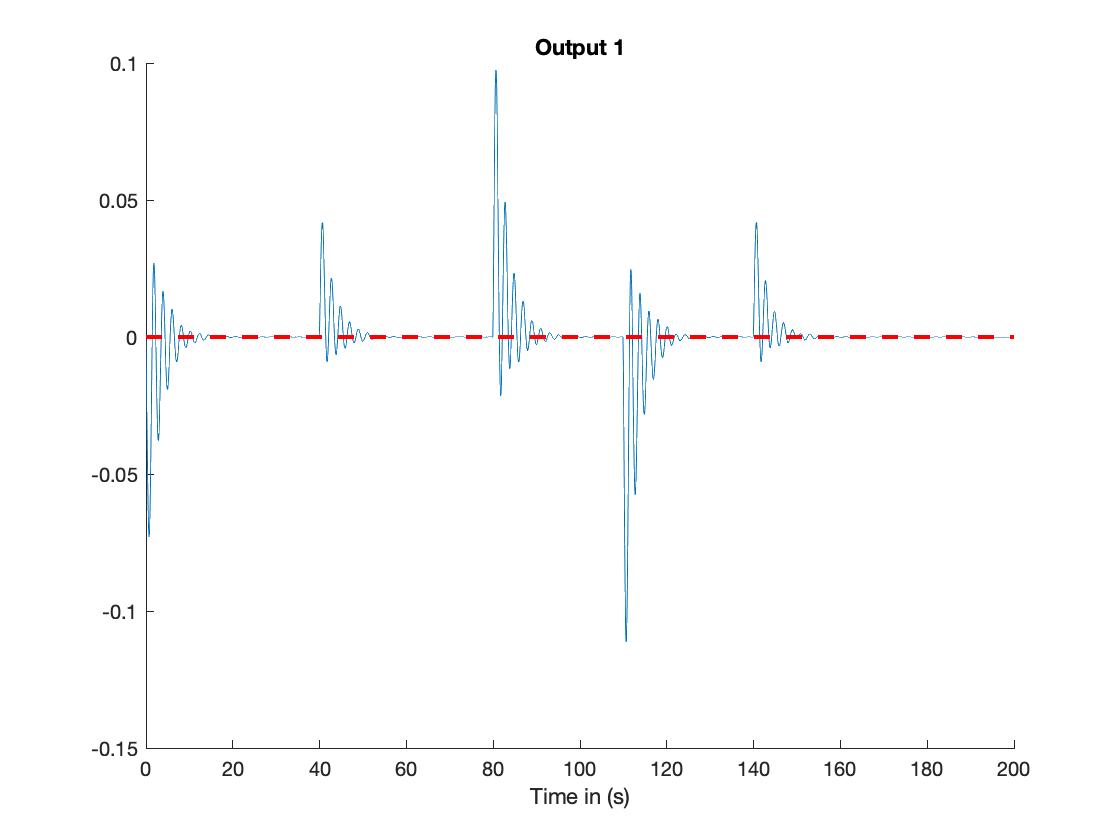,width=.29\textwidth}}
\centering%
\subfigure[\footnotesize Control line 1 ]
{\epsfig{figure=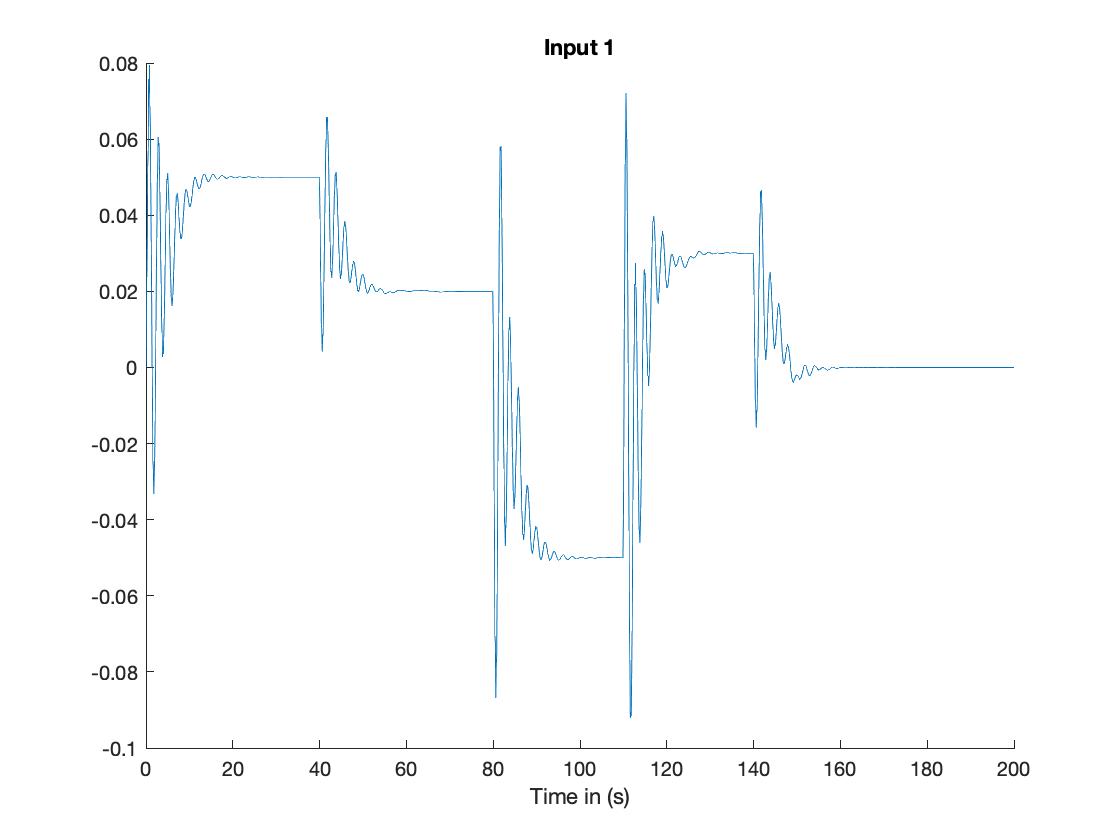,width=.29\textwidth}}
\subfigure[\footnotesize Output (--) and reference trajectory (- -) line 2 ]
{\epsfig{figure=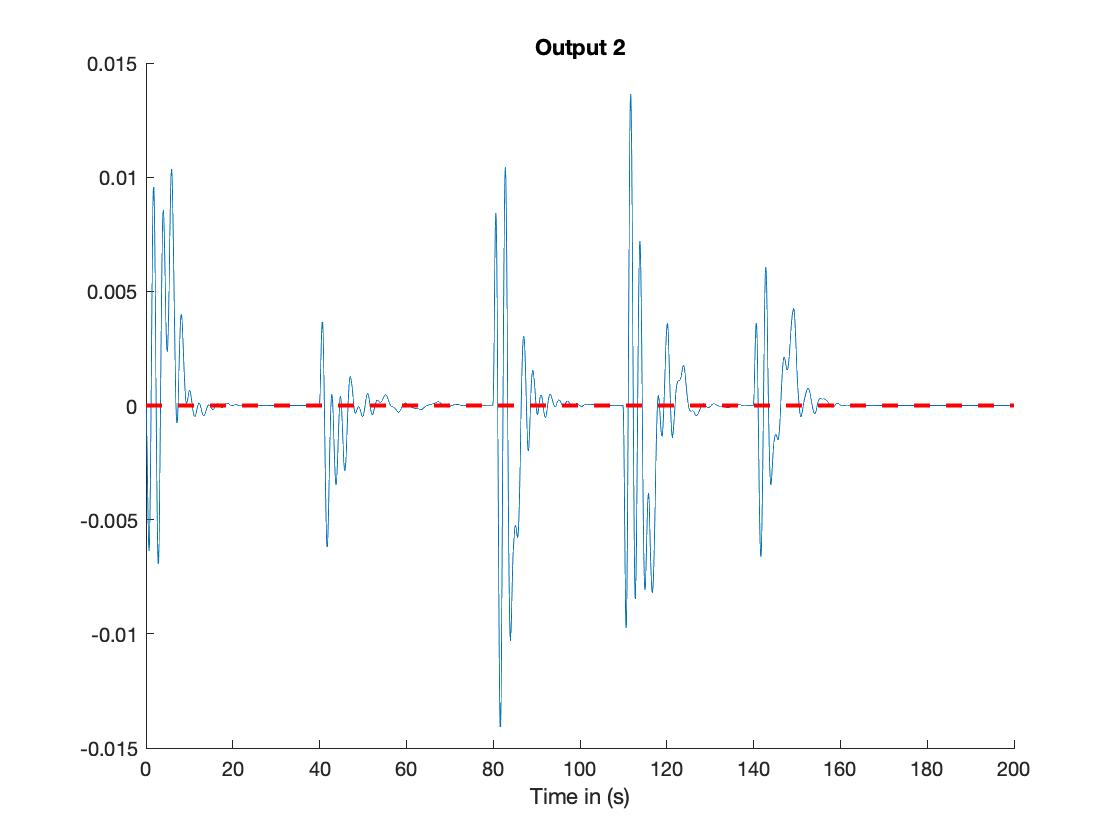,width=.29\textwidth}}
\centering%
\subfigure[\footnotesize Control line 2 ]
{\epsfig{figure=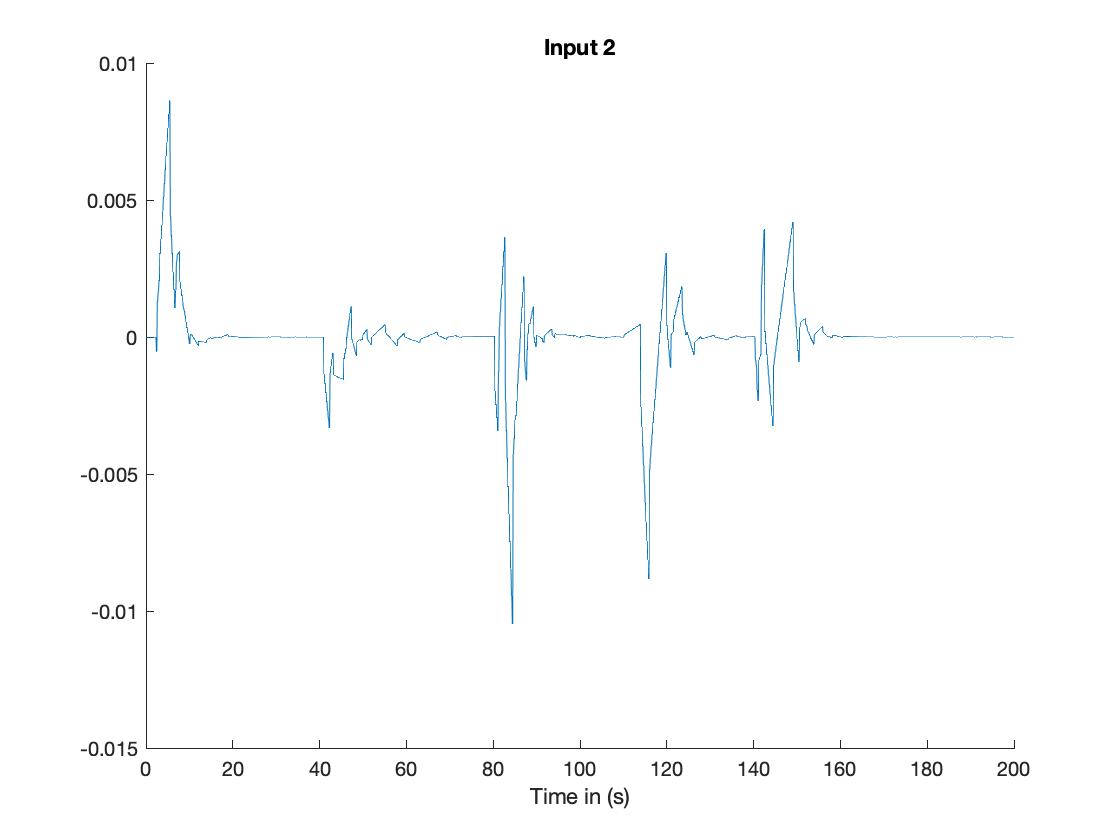,width=.29\textwidth}}
\subfigure[\footnotesize Attack-free measured output (red) and measured output after attack (blue)]
{\epsfig{figure=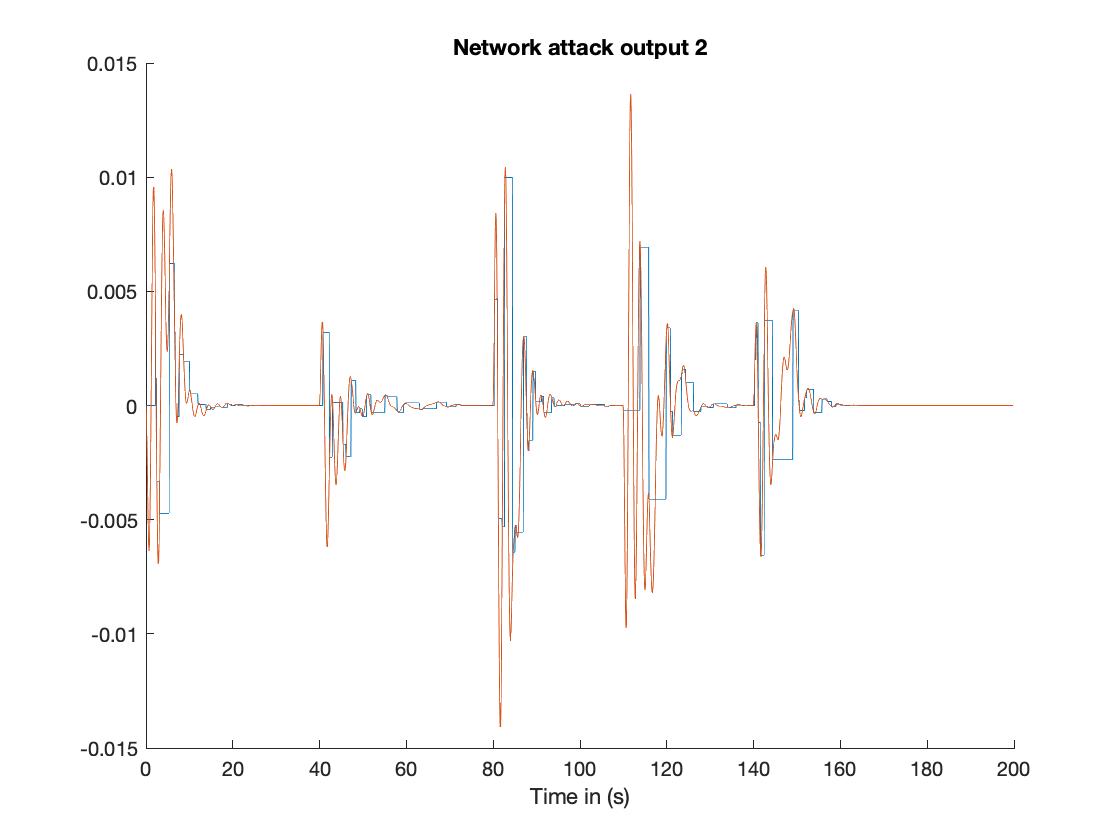,width=.29\textwidth}}
\subfigure[\footnotesize Zoom on (c)]
{\epsfig{figure=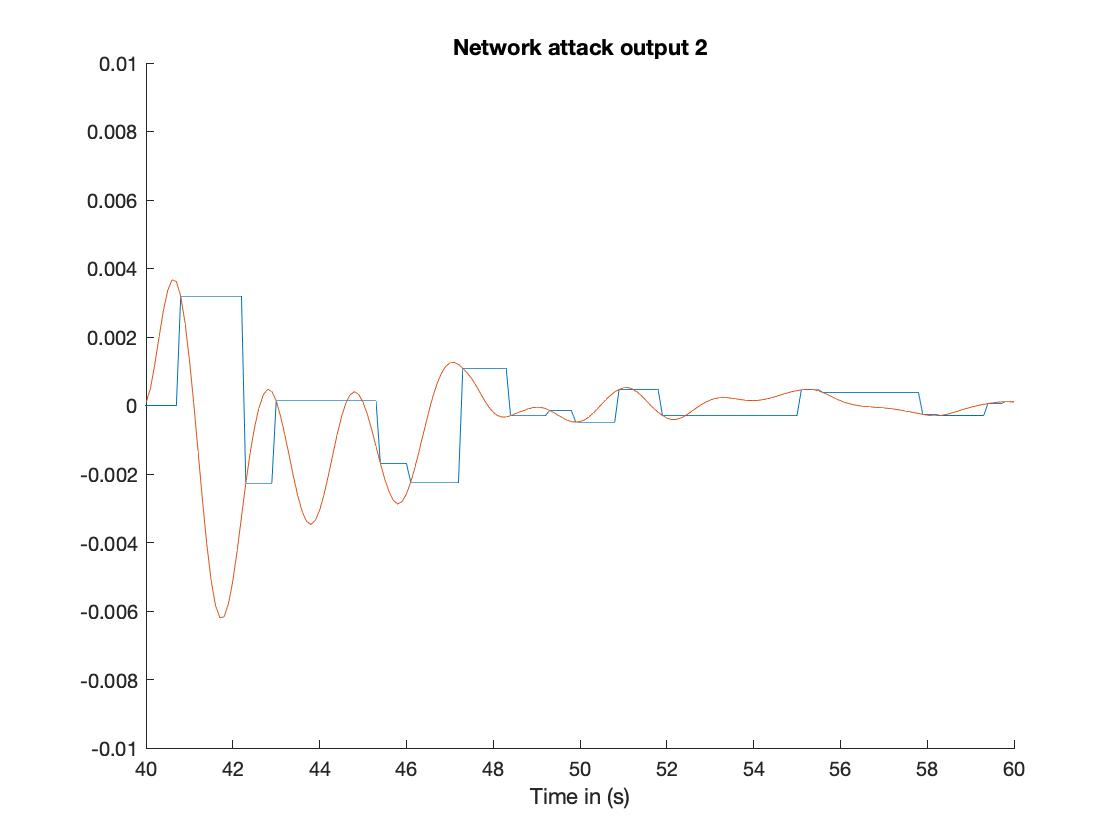,width=.29\textwidth}}
\caption{MFC: Type 3 DoS attack}\label{CSM4}
\end{figure*}

Figures \ref{PID4} and \ref{CSM4}.
\\
For each scenario, constant reference trajectory and load frequency are presented on subfigure (a) (resp. (c)) for line/area 1 (resp. 2). Subfigure (b) (Resp. (d)) draws the power control for line/area 1 (resp. 2).\\
\\
Without any attack, two control strategies seem to have similar behavior but, as shown in Table I, the trajectory tracking error decreases significantly with our proposition. \\
With type 1 attack, as illustrated in Figures \ref{PID2} and \ref{CSM2} and in Table I, better results, especially for area 2, are obtained with MFC.\\
The superiority of the model-free setting becomes crushing with in the case of DoS attacks. 
This is also highlighted by Table II: It summarizes $100$ simulations where the selection of packet loss is random and in the same proportion.


\section{Conclusion}\label{conclusion}
From a theoretical standpoint let us emphasize the two following remarks
\begin{enumerate}
\item The defense again load altering attacks is mathematically well justified in Section \ref{accom}.
\item The dazzling efficiency against DoS attacks is based only on computer experiments, \textit{i.e.}, on \emph{experimental mathematics} (see, \textit{e.g.}, \cite{arnold}). A formal proof is lacking today.
\end{enumerate}

Our proposal for cybersecurity, in order to be more convincing, needs of course further investigations, for instance on control saturation.




\mbox{}\newpage\cleardoublepage


\begin{thebibliography}{99}

\bibitem{c1}	K. Stouffer, J. Falco, K. Scarfone, Guide to industrial control systems (ics) security. NIST Spec. Public., vol. 800-82, 2007.
\bibitem{c2}	N. Falliere, Exploring stuxnet's plc infection process. http://www.symantec.com/connect/blogs/exploring-stuxnet-s-plc-infection-process, 2010.
\bibitem{c3}	I.N. Fovino,  A. Coletta, M. Masera,Taxonomy of security solutions for scada sector. JRC-Joint Research Centre Europ. Commiss., 2010.
\bibitem{c4}	R.L. Krutz (2005). Securing SCADA systems. Wiley, 2005
\bibitem{c5}	A. Cardenas, S. Amin and S. Sastry, Secure control: Towards survivable cyber-physical systems. 1st Int. Worksh. Cyber-Phys. Syst., Beijing, 2008.
\bibitem{c6}	F. Pasqualetti, Secure Control Systems: A Control-Theoretic Approach to Cyber-Physical Security. PhD, University of California, 2012.
\bibitem{c7}	F. Pasqualetti, F Dorfler, F. Bullo, Attack detection and identification in cyber-physical systems. IEEE Trans. Automat. Contr., vol. 58, 2715-2729, 2013.
\bibitem{c8}	S. Amin, A. Cardenas, S. Sastry, Safe and secure networked control systems under denial-of-service attacks. R. Majumdar  \& P. Tabuada (Eds): Hybrid Systems: Computation and Control, 
Lect. Notes Comput. Sci. 5469, 31-45, 2009.
\bibitem{c9}	Y.L. Huang, A.A. C\'ardenas, S. Amin, Z.S. Lin, H.Y. Tsai, S. Sastry. Understanding the physical and economic consequences of attacks on control systems. Int. J. Critic. Infrastruct. Protect., vol. 2, 73-83, 2009.
\bibitem{c10}	Y. Liu, M.K. Reiter, P. Ning, False data injection attacks against state estimation in electric power grids, ACM Conf. Comput. Communicat. Secur., Chicago, 2009.
\bibitem{c11}	A. Teixeira, H. Sandberg, K.H. Johansson, Networked control system under cyber attacks with applications to power networks, Amer. Contr. Conf., Baltimore, 2010.
\bibitem{c12}	R.S. Smith, A decoupled feedback structure for covertly appropriating networked control systems. World IFAC Congr., Zurich, 2011.
\bibitem{c13}	 A. Teixeira, D. Perez, H. Sandberg, K.H. Johansson, Attack models and scenarios for networked control systems.  1st Int. Conf. High Confid. Network. Syst., Beijing, 2012.
\bibitem{c14}	F. Pasqualetti, F. Dorfler, F. Bullo, Cyber-physical security via geometric control: Distributed monitoring and malicious attacks. 51st  IEEE Conf. Decis. Contr., Maui, 2012.
\bibitem{c15}	Y. Mo, B. Sinopoli, Secure control against replay attacks, Allerton Conf. Commun. Contr. Comput., Monticello, 2010.

\bibitem{review} H.S. S\'{a}nchez, D. Rotondo, J. Quevedo, Bibliographical review on cyber attacks from a control oriented perspective. Ann. Rev. Contr., vol. 48, 103-128, 2019.

\bibitem{survey}D. Bhamare, M. Zolanvari, A. Erbad, R. Jain, K. Khan, N. Meskin, Cybersecurity for industrial control systems: A survey. Comput. Secur., vol. 89, 101677, 2020.

\bibitem{md0}
M.M. Hossain, C. Peng, Cyber-physical security for on-going smart grid initiatives: a survey. IET Cyber-Phys. Syst. Theory Appl., vol. 5, 233-244, 2020.

\bibitem{zhao}
L. Zhao, W. Li,  Co-design of dual security control and communication for nonlinear CPS under DoS attack. IEEE Access, vol. 8,  19271-19285, 2020.

\bibitem{noura2}
H. Noura, D. Theilliol, J.-C. Ponsart, D. Chamseddine,
Fault-tolerant Control Systems: Design and Practical Applications. Springer, 2009.

\bibitem{blanke}
M. Blanke, M. Kinnaert, J. Lunze, M. Staroswiecki, Diagnosis and Fault-Tolerant Control (3rd ed.). Springer, 2016.

\bibitem{mfc13}
M. Fliess, C. Join, Model-free control., Int. J. Contr., vol. 86, 2228-2252, 2013.

\bibitem{nicu}
M. Fliess, C. Join, An alternative to proportional-integral and proportional-integral-derivative regulators: Intelligent proportional-derivative regulators. Int. J. Robust Nonlinear Contr., 2021. \newline {\tt \scriptsize https://doi.org/10.1002/rnc.5657}


\bibitem{wang}
Z. Wang, J. Wang, Ultra-local model predictive control: A model-free approach and its application on automated vehicle trajectory tracking. Contr. Engin. Pract., vol. 101, 104482, 2020.
\bibitem{toulon}
F. Lafont, J.-F. Balmat, N. Pessel, M. Fliess, A model-free control strategy for an experimental greenhouse with an application to fault accommodation. Comput. Electron. Agricult., vol. 110, 139-149, 2015.

\bibitem{park}
B. Park, M.M. Olama, A model-free voltage control approach to mitigate motor stalling and FIDVR for smart grids. IEEE Trans. Smart Grid, vol. 12, 67-78, 2021.

\bibitem{c25}	
H.H. Alhelou, M.E. Hamedani-Golshan, N.D. Hatziargyriou, A decentralized functional observer based optimal LFC considering unknown inputs, uncertainties, and cyber-attacks. IEEE Trans. Power Syst., vol. 34, 4408-4417, 2019.
\bibitem{c26}	
X. Liu, Y. Zhang, Y., K.Y. Lee, Robust distributed MPC for load frequencycontrol of uncertain power systems. Contr. Engin. Pract., vol. 56, 23-47, 2016.
\bibitem{c27}	S. Saxena, Y.V. Hote, Stabilization of perturbed system via IMC: An application to load frequency control. Contr. Engin. Pract., vol. 64, 61-73, 2017.
\bibitem{c28}	J. Liu, Y. Gu, L. Zha, J. Cao, Event-triggered load frequency control formultiarea power systems under hybrid cyber attacks. IEEE Trans. Syst. Man Cybernet.: Syst., vol. 49, 1665-1678, 2019..

\bibitem{md}
M.M. Hossain, C. Peng, Observer-based event triggering $H_\infty$ LFC for multi-area power systems under DoS attacks. Informat. Sci., vol. 543, 437-453, 2021. 

\bibitem{kemal}
M.S. Kemal, W. Aoudi, R.L. Olsen, M. Almgren, H.-P. Schwefel, Model-free detection of cyberattacks on voltage control in distribution grids. 15th Europ. Depend. Comput. Conf., Naples, 2015.

\bibitem{cui}
C. Chen, M. Cui, X. Fang, B. Ren, Y. Chen, Load altering attack-tolerant defense strategy for load frequency control system. Appli. Energ., vol. 280, 116015, 2020.

\bibitem{qiu} 
X. Qiu, Y. Wang, X. Xie, H. Zhang, Resilient model-free adaptive control for cyber-physical systems against jamming attack. Neurocomput., vol. 413, 422-430, 2020. 








\bibitem{loss}
H. Lin, H. Su, P. Shi, Z. Shu, Z.-G. Wu, Estimation and Control for Networked Systems with Packet Losses without Acknowledgement. Springer, 2017.


\bibitem{iot}
C. Join, M. Fliess, F. Chaxel, Model-free control as a service in the Industrial Internet of Things: Packet loss and latency issues via preliminary experiments. 28th Medit. Conf. Contr. Automat., Saint-Rapha\"{e}l, 2020. 
{\tt \scriptsize https://hal.archives-ouvertes.fr/hal-02546750/en/}

\bibitem{ex1} 
T.C. Yanga, Z.T. Ding, H. Yu, Decentralised power system load frequency control beyond the limit of diagonal dominance, Electric. Power Energy Syst., vol. 24, 173-184, 2002.
\bibitem{ex2} 
Y. Lia, P. Zhang, L. Ma, Denial of service attack and defense method on load frequency control system, J. Franklin Instit., vol. 356, 8625-8645, 2019.

\bibitem{bevrani}
H. Bevrani, Robust Power System Frequency Control (2nd ed.). Springer, 2014.



\bibitem{astrom}
K.J. {\AA}str\"om, R.M. Murray, Feedback Systems: An Introduction for Scientists and Engineers. Princeton University Press, 2008.

\bibitem{pmsm}
Y. Wang, H. Li, R. Liu, L. Yang, X. Wang, X., Modulated model- free predictive control with minimum switching losses for PMSM drive system, IEEE Access, vol. 8, 20942-20953, 2020.


\bibitem{c16}	P. Kundur, Power system stability and control. McGraw Hill, 1994.
\bibitem{c17}	J. Sharma, Y. V. Hote, R. Prasad, PID controller design for interval load frequency control system with communication time delay. Contr. Engin. Pract., vol. 9, 154-168, 2019.
\bibitem{c18}	L.  Jiang, W. Yao, Q. H. Wu, J. Y.,Wen, S. J. Cheng, Delay-dependent stability for load frequency control with constant and time-varying delays. IEEE Trans. Power Syst., vol. 27, 932-941, 2012.
\bibitem{c19}	A. Khodabakhshian, M. Edrisi, A new robust PID load frequency controller. Contr. Engin. Pract., vol. 16, 1069-1080, 2008.
\bibitem{c20}	H. H. Alhelou, M. E. Hamedani-Golshan, R. Zamani, E. Heydarian-Forushani, P. Siano, Challenges and opportunities of load frequency control in conventional, modern and future smart power systems: A comprehensive review. Energies, vol. 11, 2497, 2018.
\bibitem{c21}	S. Mishra, K. Anderson, B. Miller, K. Boyer, A. Warren,  Microgrid resilience: A holistic approach for assessing threats, identifying vulnerabilities, and designing corresponding mitigation strategies. Appl. Energy, vol. 264, 114726, 2020.
\bibitem{c22}	M. Z.  Gunduz, R. Das, Cyber-security on smart grid: Threats and potential solutions. Comput. Netw., vol. 169, 107094, 2020 
\bibitem{c23}	 D. Zhang, Q.G. Wang, G.F. YangShi, A.V. Vasilakos,  A survey on attack detection, estimation and control of industrial cyber-physical systems. ISA Trans., 2021
\bibitem{c24}	D.B. Rawat, C. Bajracharya, Detection of false data injection attacks in smart grid communication systems. IEEE Signal Process Lett, vol. 22, 1652-1654, 2015.
\bibitem{c29}	H. Noura, D. Sauter, F. Hamelin, D. Theilliol, Fault-tolerant control in dynamic systems: Application to a winding machine. IEEE Contr. Syst. Magaz., vol. 20, 33-49, 2000.
\bibitem{c30}	C. Chen , M. Cui , X. Fang , B. Ren, Y. Chen, Load altering attack-tolerant defense strategy for load frequency control system. Appl. Energy, vol. 280, 2020.
\bibitem{c31}	X. Qiu, Y. Wang, X. Xie,  H. Zhang  Resilient model-free adaptive control for cyber-physical systems against jamming attack. Neurocomput., vol. 413, 422-430, 2020. 

\bibitem{sarajevo}
A. Barkat, B. Marinescu, C. Join, M. Fliess, Model-free control for VSC-based HVDC systems. IEEE PES Innovat. Smart Grid Techno. Conf. Euro., Sarajevo, 2018. \newline
{\tt \scriptsize https://hal.archives-ouvertes.fr/hal-01820886/en/}

\bibitem{ferrari}
M. Ferrari, B. Park, M.M. Olama, Design and evaluation of a model-free frequency control strategy in islanded microgrids with power-hardware-in-the-loop testing. 
IEEE Pow. Ener. Soc. Innov. Smart Grid Tech. Conf., Washington, 2021.

\bibitem{alinea}
H. Aboua\"{\i}ssa, M. Fliess, C. Join, On ramp metering: towards a better understanding of ALINEA via model-free control. Int. J. Contr., vol. 90, 1018-1026, 2017.

\bibitem{arnold}
V.I. Arnold,
Experimental Mathematics (translated from the Russian),
Math. Sci. Res. Instit., 2015.

\end{thebibliography}
\end{document}